\documentclass[longauth]{aa}  
\usepackage{graphicx}                  
\usepackage[varg]{txfonts}             
\usepackage[dvipsnames]{xcolor}
\usepackage{hyperref}
\hypersetup{pdfborder = {0 0 0},
    colorlinks = true,
    linkbordercolor = {white},
    citecolor=NavyBlue,
    urlcolor=Green
}

\newcommand{\orcidlink}[1]{\protect\href{https://orcid.org/#1}{\protect\includegraphics[width=8pt]{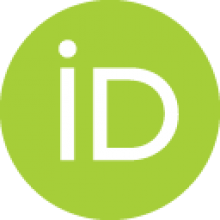}}}

\usepackage{amssymb}
\usepackage{amsmath}
\usepackage{subcaption}
\usepackage{xspace}
\usepackage{array,multirow}
\usepackage{enumitem}
\usepackage[english]{babel}
\usepackage{mathrsfs}
\usepackage{listings}
\usepackage{blindtext}
\usepackage{lipsum} 
\usepackage{booktabs}
\usepackage{url}
\usepackage{natbib,twoopt}
\usepackage{placeins}
\usepackage[normalem]{ulem}  
\usepackage{CJKutf8}

\newcommand{\ie}{i.e.\@\xspace} 
\newcommand{\eg}{e.g.\@\xspace} 

\renewcommand{\arraystretch}{1.}

\renewcommand{\eqref}[1]{Eq.~\ref{#1}}
\newcommand{\fref}[1]{Fig.~\ref{#1}}

\newcommand{\tref}[1]{Table~\ref{#1}}
\newcommand{\sref}[1]{Sect.~\ref{#1}}

\newcommand{\aref}[1]{Appendix~\ref{#1}}

\newcommand{\code}[1]{{\texttt{#1}}}

\newcommand{\CNnames}[1]{{\begin{CJK}{UTF8}{gbsn}(#1)\end{CJK}}}

\newcommand{\numax}{\ensuremath{\nu_{\rm max}}\xspace}
\newcommand{\dnu}{\ensuremath{\Delta\nu}\xspace}

\newcommand{\msol}{\ensuremath{\rm M_{\odot}}\xspace}

\newcommand{\teff}{\ensuremath{T_{\rm eff}}\xspace}

\newcommand{\vsini}{\ensuremath{v\sin i}\xspace}

\numberwithin{equation}{section}

\makeatletter
\def\maketag@@@#1{\hbox{\m@th\normalfont\normalsize#1}}
\makeatother

\bibpunct{(}{)}{;}{a}{}{,} 

\makeatletter
\newcommand*\mysize{%
  \@setfontsize\mysize{5.7}{8.0}%
}
\makeatother

\makeatletter
\newcommand*\tabsize{%
  \@setfontsize\tabsize{7.}{8.0}%
}
\makeatother

\makeatletter
\newcommand\footnoteref[1]{\protected@xdef\@thefnmark{\ref{#1}}\@footnotemark}
\makeatother

\begin{document}

   \title{The Stellar Observations Network Group (SONG)}

   \subtitle{A legacy archive of stellar time-domain spectroscopy}

    \author{M. N. Lund\inst{\ref{DK}}\corrauth{mikkelnl@phys.au.dk}\orcidlink{0000-0001-9214-5642} \and 
   F. Grundahl\inst{\ref{DK}}\email{fgj@phys.au.dk}\orcidlink{0000-0002-8736-1639} \and
   M. S. Fredslund\inst{\ref{DK}}\email{madsfa@phys.au.dk}\orcidlink{0000-0002-9194-8520} \and
P.~L.~Pall\'{e}\inst{\ref{IAC1},\ref{IAC2}}\email{pere.l.palle@iac.es}\orcidlink{0000-0003-3803-4823} \and 
   S. Simon-Diaz\inst{\ref{IAC1},\ref{IAC2}}\email{sergio.simon.diaz@iac.es}\orcidlink{0000-0003-1168-3524} \and 
   J.~Christensen-Dalsgaard\inst{\ref{DK}}\email{jcd@phys.au.dk}\orcidlink{0000-0001-5137-0966}   \and 
   R.~A.~Wittenmyer\inst{\ref{aus}}\email{Rob.Wittenmyer@unisq.edu.au}\orcidlink{0000-0001-9957-9304} 
   \and L.~Deng \CNnames{邓李才}\inst{\ref{China1},\ref{China2}}\email{licai@nao.cas.cn}\orcidlink{0000-0001-9073-9914} \and 
   J.~Jackiewicz\inst{\ref{NMS}}\email{jasonj@nmsu.edu}\orcidlink{0000-0001-9659-7486}\and
   H.~Kjeldsen\inst{\ref{DK}}\email{hans@phys.au.dk}\orcidlink{0000-0002-9037-0018} \and  
   V.~Antoci\inst{\ref{dtu}}\email{antoci@space.dtu.dk}\orcidlink{0000-0002-0865-3650} \and 
    H.~Korhonen\inst{\ref{Heidi1}}\email{korhonen@mpia.de}\orcidlink{0000-0003-0529-1161}\and
    S.~H.~Albrecht \inst{\ref{DK}}\email{albrecht@phys.au.dk}\orcidlink{0000-0003-1762-8235} \and 
   K.~Wang \CNnames{王坤}\inst{\ref{China2}}\email{kwang@bao.ac.cn}\orcidlink{0000-0002-5745-827X} \and    
J.~Rudrasingam\inst{\ref{DK},\ref{Tim}}\email{jrud0912@uni.sydney.edu.au}\orcidlink{0009-0007-7973-9228} \and
 R.~Handberg\inst{\ref{DK}}\orcidlink{0000-0001-8725-4502} \and
   T.~R.~Bedding\inst{\ref{Tim}}\email{tim.bedding@sydney.edu.au}\orcidlink{0000-0001-5222-4661} \and 
   J.~L.~R\o rsted\inst{\ref{DK}}\email{jakob@phys.au.dk}\orcidlink{0000-0001-9234-430X} \and  
   P.~M.~S\o rensen\inst{\ref{NOT},\ref{DK}}\email{pms@phys.au.dk} \and
   D.~J.~Wright\inst{\ref{aus}}\email{Duncan.Wright@unisq.edu.au}\orcidlink{0000-0001-7294-5386} \and
   J.~A.~Holtzman\inst{\ref{NMS}}\email{jholtzma@nmsu.edu}\orcidlink{0000-0002-9771-9622} \and 
  J.~Klusmeyer\inst{\ref{NMS}}\email{jaklus@nmsu.edu}\orcidlink{0000-0003-3906-9518} \and 
   S.~Frandsen\inst{\ref{DK}}\email{srf@phys.au.dk} \and
    E.~Weiss\inst{\ref{DK}}\email{eric@eric-weiss.de} \and
    J.~Jessen-Hansen\inst{\ref{DK}}\email{jensjessenhansen@gmail.com}\orcidlink{0000-0003-4985-7606} \and 
    P.~Heeren\email{ppheeren@posteo.de}\orcidlink{0000-0002-3662-9930} \and 
R.~Tronsgaard\inst{\ref{DK},\ref{cam}}\email{rtr@phys.au.dk}\orcidlink{0000-0003-1001-0707} \and 
  T.~Arentoft\inst{\ref{DK}}\email{toar@au.dk}\orcidlink{0000-0002-4696-6041} \and 
P.~G.~Beck\inst{\ref{IAC1},\ref{IAC2}}\email{paul.beck@iac.es}\orcidlink{0000-0003-4745-2242} \and 
  U.~G.~J\o rgensen\inst{\ref{nbi1}}\email{uffegj@nbi.ku.dk}\orcidlink{0000-0001-7303-914X} \and 
  A.~N.~S\o rensen\inst{\ref{nbi1}}\email{norup@nbi.ku.dk}\orcidlink{0000-0002-7106-2781} \and M.~I.~Andersen\inst{\ref{nbi1}}\email{manderse@nbi.ku.dk}\orcidlink{0000-0002-8109-033X} \and 
  P.~Kjærgaard\inst{\ref{nbi1}}\email{per@nbi.ku.dk} \and 
  M.~Bizzarro\inst{\ref{globe}}\email{bizzarro@sund.ku.dk}\orcidlink{0000-0001-9966-2124} \and
  R.~P.~Butler\inst{\ref{carnigie}}\email{paul@dtm.ciw.edu}\orcidlink{0000-0003-1305-3761} \and
  E.~Corsaro\inst{\ref{INAF}}\email{enrico.corsaro@inaf.it}\orcidlink{0000-0001-8835-2075} \and 
  R.~A.~Garc\'{i}a\inst{\ref{paris}}\email{rafael.garcia@cea.fr}\orcidlink{0000-0002-8854-3776} \and
D.~Stello\inst{\ref{dennis1},\ref{Tim}}\email{d.stello@unsw.edu.au}\orcidlink{0000-0002-4879-3519} \and 
S.~Murphy\inst{\ref{aus}}\email{Simon.Murphy@unisq.edu.au}\orcidlink{0000-0002-5648-3107} \and 
J.~Horner\inst{\ref{aus}}\email{jonti.horner@unisq.edu.au}\orcidlink{0000-0002-1160-7970} \and
S.~Martell\inst{\ref{dennis1}}\email{s.martell@unsw.edu.au}\orcidlink{0000-0002-3430-4163} \and
M.~J.~Martínez~González\inst{\ref{IAC1},\ref{IAC2}}\email{m.j.martinez@iac.es}\orcidlink{0000-0001-5560-7502} \and L. Casagrande\inst{\ref{luca}}\email{luca.casagrande@anu.edu.au}\orcidlink{0000-0003-2688-7511} \and
J. Lattanzio\inst{\ref{monash}}\email{john.lattanzio@monash.edu}\orcidlink{0000-0003-2952-859X}}
\institute{
    Stellar Astrophysics Centre (SAC), Department of Physics and Astronomy, Aarhus University, Ny Munkegade 120, 8000 Aarhus C, Denmark\label{DK} \and
Instituto de Astrofísica de Canarias, E-38200 La Laguna, Tenerife, Spain\label{IAC1} \and
Departamento de Astrofísica, Universidad de La Laguna (ULL), 38206 La Laguna, Tenerife, Spain\label{IAC2} \and
Centre for Astrophysics, University of Southern Queensland, Toowoomba, QLD 4350, Australia\label{aus} \and
National Astronomical Observatories, Chinese Academy of Sciences, Beijing 100101, China\label{China1} \and
School of Physics and Astronomy, China West Normal University, Nanchong 637009, People’s Republic of China\label{China2} \and
New Mexico State University, Department of Astronomy, MSC 4500, Box 30001, Las Cruces NM 88011\label{NMS} \and DTU Space, Technical University of Denmark, Elektrovej 327, Kgs. Lyngby 2800, Denmark \label{dtu} \and
Max-Planck-Institut für Astronomie, Königstuhl 17, DE-69117, Heidelberg, Germany\label{Heidi1} \and 
Sydney Institute for Astronomy, School of Physics, University of Sydney, Sydney, NSW 2006, Australia\label{Tim} \and
Nordic Optical Telescope (NOT), Santa Cruz de la Palma, Canary Islands, Spain \label{NOT}\and
Affiliate Member, Cavendish Laboratory, University of Cambridge, Cambridge CB3 0US, UK\label{cam}\and
Centre for ExoLife Sciences, Niels Bohr Institute, University of Copenhagen, Jagtvej 155, 2200 Copenhagen, Denmark \label{nbi1} \and
Centre for Star and Planet Formation, GLOBE Institute, University of Copenhagen, Øster Voldgade 5-7, 1350 Copenhagen, Denmark\label{globe} \and
Earth and Planets Laboratory, Carnegie Institution for Science, 5241 Broad Branch Road, NW, Washington, DC 20015, USA\label{carnigie} \and
INAF – Osservatorio Astrofisico di Catania, Via S. Sofia, 78, 95123 Catania, Italy\label{INAF} \and
Université Paris-Saclay, Université Paris Cité, CEA, CNRS, AIM, 91191, Gif-sur-Yvette, France\label{paris} \and
School of Physics, University of New South Wales, NSW 2052, Australia\label{dennis1} \and Research School of Astronomy and Astrophysics, The Australian National University, Canberra, ACT 2611, Australia\label{luca} \and School of Physics \& Astronomy, Monash University, Wellington Road, Clayton 3800, Victoria, Australia\label{monash}
    }
    \authorrunning{Lund et al.}
   \date{Received June 22, 2026 / accepted August 14, 2026}
 
  \abstract
   {The Stellar Observations Network Group (SONG) network has operated for more than a decade, providing long-baseline high-cadence spectroscopic observations of bright stars and the Sun. The observations from 2014 through 2025 constitute a substantial archive of high-resolution spectra and precise radial velocities suitable for a broad range of time-domain stellar astrophysics.}
   {We present an overview of the current status, instrumentation, and scientific capabilities of the SONG network and describe the scope and accessibility of the SONG data archive (SODA). We further illustrate the breadth of science enabled by SONG observations, including asteroseismology, stellar variability studies, binary-star characterisation, and exoplanet research.}
   {We summarise the operational status and observing strategies of the SONG facilities, describe the available data products and archive infrastructure, and outline procedures for accessing archival observations and proposing new observations within the SONG community framework.}
   {The SODA archive currently contains more than 580,\,000 spectra of 3091 stars obtained with SONG using either iodine-cell or thorium–argon wavelength calibration. The archive spans more than a decade of observations and includes extensive high-cadence time-series data for bright targets spanning a wide range of stellar types and variability classes. Access to the archive is available to members of the SONG community, which remains open to new participants who agree to follow the community policies.}
   {The SONG archive has developed into a major long-baseline resource for stellar spectroscopy and radial velocity time-series analysis. Continued expansion of the archive, together with coordinated observations obtained contemporaneously with TESS and future PLATO observations, is expected to enable new studies of stellar oscillations, variability, and exoplanet host stars through combined radial velocity and photometric analyses.}

   \maketitle
   \nolinenumbers
%
\section{Introduction}
\begin{figure*}
   \includegraphics[width=\textwidth]{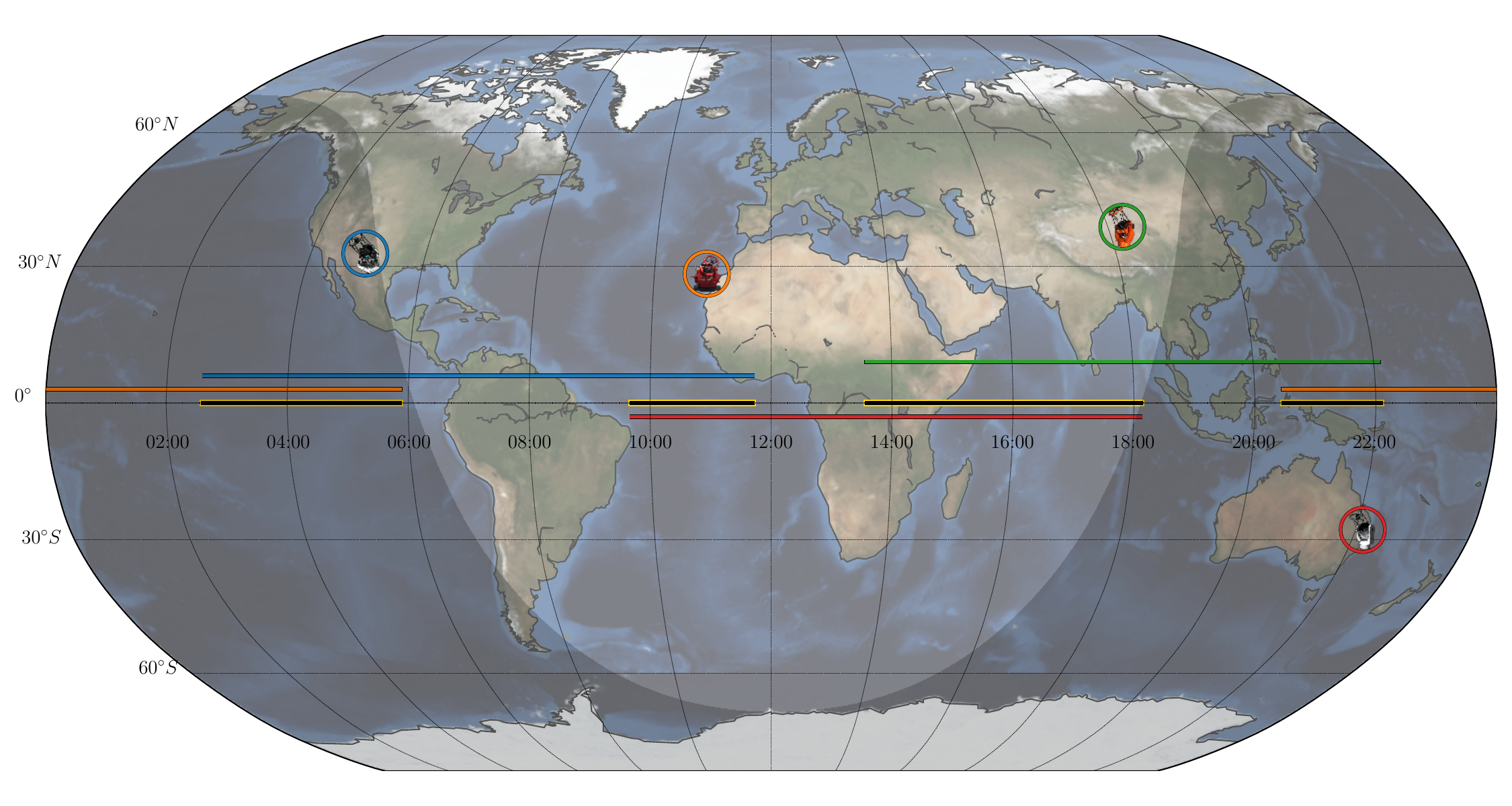}
      \caption{Overview of SONG sites, with the geographical location of the four SONG sites currently in commissioning or in operation indicated (Robinson projection). OT is encircled in orange, MtK in red, Lenghu in green, and APO in blue (see \tref{tab:sites}). With colours corresponding to the different sites, the horizontal lines around the equator (ordered by site in latitude) show the site's observing night in UT (not longitude; see labels at $-10^{\circ}$ latitude) for the star $\beta$ Aql on July 20, 2025, with the criteria of $>20^{\circ}$ altitude and the Sun $<-6^{\circ}$ below the horizon. In the current example, the four sites enable 24-hour observing coverage, and the equatorial gold and black lines indicate the periods when two sites observe $\beta$ Aql contemporaneously. We also show the day-to-night terminator (civil twilight) at noon UT on the same day. }  
         \label{fig:sites}
   \end{figure*}

Asteroseismology of solar-like stars based on radial velocity (RV) measurements demands near-continuous high-precision time series. Daily gaps in observations introduce spectral window artefacts that can mimic or mask real oscillation frequencies, fundamentally limiting the precision of mode identification and frequency measurements \citep{Arentoft2014}. The Stellar Observations Network Group (SONG) was designed to meet this challenge through a global network of 1 m telescopes that can operate as a single unified instrument \citep{Grundahl2007, Andersen2019b}. The primary science driver is asteroseismology of solar-like stars \citep{Grundahl2017, Malla2020, Knudstrup2023, Kjeldsen2025}, with the network's geographic distribution (\fref{fig:sites}) enabling the required near-continuous coverage. Beyond asteroseismology, the high-resolution spectroscopic capability of SONG naturally supports a broader range of stellar science, including the discovery and characterisation of exoplanets and their host stars \citep{Addison2021, Subjak2023}, long-term monitoring of stellar variability \citep{Korhonen2021, Cao2022, SimonDiaz2018, SimonDiaz2024}, the study of binary systems \citep{Brogaard2021, Hey2022}, and the general determination of stellar parameters \citep{Henriksen2023}.

The first node of the network, the Hertzsprung SONG Telescope \citep{Andersen2016}, was inaugurated in 2014 at the Observatorio del Teide (OT) in Tenerife, Spain, through a collaboration between the Department of Physics and Astronomy at Aarhus University (AU) and the Instituto de Astrofísica de Canarias (IAC; \sref{sec:OT}). In 2014, a second node was also established at the Delingha Observatory in China \citep{SONG_China2018}, operated by the National Astronomical Observatories of the Chinese Academy of Sciences (NAOC), and it was subsequently relocated to the newly established Lenghu Observatory in Qinghai province \citep{Deng2021}, where commissioning is ongoing (\sref{sec:lenghu}). A third node became operational in 2020 at the Mt. Kent (MtK) Observatory in Australia \citep{Holt2022}, in partnership with the University of Southern Queensland, University of New South Wales, Monash University, The Australian National University, and the University of Sydney (\sref{sec:mtk}). Most recently, a fourth node is being commissioned at Apache Point Observatory (APO) in New Mexico, USA, in collaboration with New Mexico State University (\sref{sec:apo}). 

A central aim of this paper is to document the observational legacy accumulated by SONG since its inception in 2014, comprising some $580,000$ spectra of 3091 stellar targets, and to present the framework for broader community access and participation through the SONG Data Archive (SODA)\footnote{\label{sodaurl}\url{https://soda.phys.au.dk/}}. Unlike space-based photometric missions such as the Transiting Exoplanet Survey Satellite \citep[TESS;][]{Ricker2015} and the forthcoming PLAnetary Transits and Oscillations of stars \citep[PLATO;][]{Rauer2025} mission, SONG provides ground-based high-resolution spectroscopy and precise RVs, offering a complementary observational capability that is particularly powerful when the two approaches are combined. With recent efforts to adopt a more open and collaborative model, we aim to expand the SONG community and encourage wider use of its extensive observational dataset for archival studies and new observing programs.

The paper is structured as follows. Section \ref{sec:sites} describes the network sites and instrumentation. \sref{sec:extract} covers data extraction and available data products. Section \ref{sec:obs} outlines observing capabilities and strategy, and \sref{sec:overview} presents a statistical overview of observations obtained between 2014 and 2025. Section \ref{sec:science} highlights the primary science drivers with examples of studies conducted to date. Section \ref{sec:use} describes access to the SONG data and how new observations can be proposed. We conclude in \sref{sec:con} with a summary and future outlook.

\section{SONG sites and instrumentation}\label{sec:sites}

The original concept for SONG was to establish a prototype node with a 1 m telescope and a high-resolution spectrograph, followed by adding nodes to complete the network. 
Initially, the aim was a network of 70 cm telescopes with a single-focus platform, but increased funding made it possible to expand the ambition to a more powerful design of 1 m telescopes with a Nasmyth platform on each side of the telescope. This allows us to acquire data faster and to add more instruments.

Due to real-world constraints on funding, manpower, and site-specific factors, the sites are similar but not identical in their instrumental setups. The characteristics of each SONG site and its instrumentation are provided in \tref{tab:sites}, and the site's geographical distribution is provided in \fref{fig:sites}.

\begin{table*}
    \centering
    \caption{Site and instrumentation details for the SONG nodes.}
    \label{tab:sites}
    \renewcommand{\arraystretch}{1.1} 
    \begin{tabular}{@{\extracolsep{\fill}}lcccccc@{}}
    \toprule
                      & {\bf OT}     & {\bf Lenghu} & {\bf MtK} & {\bf APO}    \\
    \midrule  [0.3ex]
    \multicolumn{5}{c}{Site information} \\ 
    \midrule
    Latitude     & $28^\circ 18' 1.66'' \, \mathrm{N}$  & $38^\circ 36' 24.48'' \, \mathrm{N}$  & $27^\circ 47' 52.38'' \, \mathrm{S}$  &   $32^\circ 46' 48.15'' \, \mathrm{N}$ \\
    Longitude    & $16^\circ 30' 44.58'' \, \mathrm{W}$ & $93^\circ 53' 45.96'' \, \mathrm{E}$ &  $151^\circ 51' 20.19'' \, \mathrm{E}$ & $105^\circ 49' 13.10'' \, \mathrm{W}$ \\
    Altitude (m)       & 2400        & 4500         & 700               & 2850 \\
    Usable nights (\%) & $85\pm5$ & TBE & $45\pm5$ & TBE \\
    \midrule [0.3ex]
    \multicolumn{5}{c}{Instrumentation information} \\ 
    \midrule 
    Telescope aperture & 1.0 m        & 1.0 m        & 2$\times$0.7 m     & 1.0 m     \\
    Focal ratio        & 41           & 36           & 6.5               & 6.5       \\
    Manufacturer       & Astelco      & NIAOT        & PlaneWave                & PlaneWave        \\
    Installation year  & 2012         & 2020         & 2020               & 2025      \\
    Spectral focus        & Coud{\'e}    &  Coud{\'e}   & Fibre              & Fibre     \\
    Resolution         & 130000       & 110000       & 100000             & 100000    \\
    Wavelength range ({\AA})   & 4400-6700    &  4400-6700   & 4400-6700         & 4400-6700 \\
    Detector  (--2024)  & Andor IKon-L & Andor-IKon-L & Kepler4040 &     \\
    Detector (2024-- ) & QHY600       & QHY600       & QHY600     & QHY600 \\
    \bottomrule
    \end{tabular}   
    \tablefoot{TBE means `to be estimated' (the usable nights fraction for actual SONG observations will be estimated for Lenghu and APO when these nodes have conducted observations for at least a full year). All sites include an iodine cell.}
\end{table*}

\subsection{Spectrograph commonalities}

Each SONG site has a standard high-resolution cross-dispersed {\'e}chelle spectrograph as its primary instrument, and the optical design is the same. An R4 Newport {\'e}chelle grating is used in each case, with 31.6 lines per millimetre, dimensions of 90 by 340 mm, and a thickness of 50 mm. 
It is worth noting that as we adopted the iodine-cell method for measuring RVs, the demands on environmental control are looser than for other spectrographs such as the Echelle SPectrograph for Rocky Exoplanets and Stable Spectroscopic Observations \citep[ESPRESSO;][]{Pepe2021}, the NN-EXPLORE Exoplanet Investigations with Doppler spectroscopy \citep[NEID;][]{Schwab2016}, or the High Accuracy Radial Velocity Planet Searcher \citep[HARPS;][]{Mayor2003}, which helps to reduce costs. 

At all the nodes, we can pass the telescope beam through an iodine cell. However, there are slight differences in the placement of the cell in the optical path. At MtK, the iodine cell is made with wedged counter-rotated end windows. This alleviates fringing when the cell is located in a collimated beam. For the OT and Delingha and Lenghu sites, the cell end windows are plane-parallel and thus produce low-amplitude fringing in the spectrum when placed in a collimated beam. For the APO node, the cell is placed in the converging beam at the telescope Nasmyth focus, thereby avoiding spectral fringing. For non-iodine observations, the spectra are not affected by fringing.

\subsection{Observatorio del Teide (OT)}\label{sec:OT}
The instrumentation at OT consists of a 1 m telescope from ASTELCO Systems\footnote{\url{https://www.astelco.com/}}, which is equipped with active optics and a coud{\'e} focus located in a standard 20-foot shipping container next to the telescope pier \citep{Andersen2014,Andersen2016}. The spectrograph is located in the container, along with all computers and power supplies for controlling the telescope and spectrograph functions. The spectrograph is enclosed in an insulated box with heating elements on each side to assist with temperature control. The air temperature in the container is controlled by standard home air-conditioning units to approximately $\rm 1.5^{\circ}C$. A slit mask with multiple positions is installed. This allows the user to select the desired spectral resolution. Since November 4 2025, however, this has been locked at the highest resolution (slit 8). At OT, the observing conditions are outstanding, with typical seeing of around 1 arcsecond. This has, in practice, led us to only use slits with widths of $1.1''$ and $0.8''$ to avoid severely under-filling the slit.  

SONG is primarily designed for time-series observations of bright targets, and the auto-guiding system is not optimised for faint targets, with $V = 9.5$ as the current limit. 
With the spectrograph located at a coud{\'e} focus, guiding functions are performed by a guide camera in the container. The auto-guide system uses a cube beamsplitter (10/90 split) to direct collimated light to the guide camera via a small wedge mirror. This setup also allows us to monitor the telescope focus, so the system continually auto-focuses during observations.

The detector we used initially was an Oxford Instruments\footnote{\url{https://andor.oxinst.com/}} Andor IKON-L CCD. From the node installation to April 2018, the wavelength range was $4382-6904\,\AA$ across 51 orders. In April 2018, the detector was rotated 90 degrees, which shifted the spectral order by 1, resulting in a slight change in the wavelength range to $4245-6828\,\AA$. This detector was not large enough to cover the full extent of the {\'e}chelle orders at wavelengths greater than $5300\,\AA$, and thus, the spectra from this detector have small wavelength gaps.
In January 2024, the detector was replaced by a QHY600 CMOS by QHYCCD\footnote{\url{https://www.qhyccd.com/}}.
The Andor detector rotation and change to the QHY detector were motivated by a problem encountered in the high-precision RV time series obtained for standard stars, where a recurring pattern with a period of one year was seen for all stars. This now appears to be largely solved (see \aref{sec:oneyear} for additional details). The QHY detector also has much smaller pixels than the IKON-L detector ($\rm 3.76\,\mu m$ vs. $\rm 13.5\,\mu m$) and lower readout noise.  

\subsection{Mt. Kent Observatory (MtK)}\label{sec:mtk}
At the MtK site, light is collected by two 70 cm CDK700 telescopes from PlaneWave Instruments\footnote{\url{https://planewave.com}} and fed to the spectrograph through octagonal optical fibres with a core diameter of $50\,\rm\mu m$. Each telescope is equipped with a guide camera and a channel that can feed calibration light into the science fibre. The spectrograph is located in a nearby building and enclosed in two insulated boxes: an inner box with 25 mm styrofoam walls, which houses the spectrograph's optical table, and a smaller optical table that contains the iodine cell and slit-viewing camera. Surrounding this is an outer box (100 mm wall). Below the optical table, two fans circulate air between the inner and outer boxes. The room temperature is controlled by the building AC system and varies by $\rm 2^{\circ}C$ peak-to-peak on a timescale of 30 minutes. Due to the two-box structure, the coupling of these temperature variations into the inner box is generally below $\rm 0.1^{\circ}C$ over long timescales.

With our current setup, the two telescopes always observe the same target. In principle, it would be possible to observe two targets simultaneously as the object traces are well separated on the detector.  The initial detector used for spectroscopic observations was the Kepler 4040 camera from Finger Lakes Instrumentation\footnote{\url{https://flicamera.com}}. This was replaced by a QHY600 detector (February 2025) because the Kepler 4040 exhibited significant residual images, non-linearity, and numerous hot pixels. 

\subsection{Lenghu Observatory}\label{sec:lenghu}
The Chinese SONG node was initially established at Delingha following the first SONG node at OT. Its 1 m telescope employs active optics and uses a coud{\'e} optical path to direct light to the spectrograph. The spectrograph system, based on Danish designs identical to the OT node, was engineered and fabricated by the Nanjing Institute of Astronomical Optical Technology (NIAOT). After successful system testing in late 2013, the facility operated manually until 2017, when light pollution from urban development necessitated its closure. The instrumentation was subsequently returned to NIAOT for refurbishment.

In 2024, the system was recommissioned at the Lenghu site, located approximately 400 km west of Delingha. Current upgrades include replacing the science camera with a QHY-600pro model, while the automation systems remain under development.
The Lenghu site has shown truly excellent observational conditions, superior to those at the Delingha site, with a median seeing of 0.75 arcseconds and over $80\%$ of nighttime hours annually usable. 

\subsection{Apache Point Observatory (APO)}\label{sec:apo}

The APO SONG node\footnote{\url{http://astronomy.nmsu.edu/song-wiki/}} had first light in October 2025 and has, at the time of writing, been collecting science data for several months on a variety of targets. This node is equipped with a PlaneWave CDK1000 telescope that is fully robotically controlled. A focal plane unit with the iodine cell and
a Shelyak FIGU (fibre injection and guiding unit) is installed at the Nasmyth focus. Calibration light sources are also integrated at the telescope focal plane and can be injected into the science fibre.

The spectrograph is installed at APO in a climate-controlled room near the Sloan Digital Sky Survey 2.5 m telescope. It is inside a ${\sim}50$ mm foam-insulated box for further environmental control. Similar to the MtK node, the spectrograph is fed with $53\,\rm\mu m$ fibres (circular, with small sections of fused octagonal), but over a length of about 80 metres.

\section{SONG data}\label{sec:extract}

\subsection{Data extraction}\label{sec:extract2}
All SONG data are collected fully automatically, from daily calibrations to nightly observations. Subsequently, all data are automatically transferred to Aarhus for further processing, which consists of spectrum extraction and RV calculation from iodine-cell data. We note that non-iodine data, that is, data for which the wavelength solution is based on thorium-argon (ThAr) spectra, do not currently have an automated RV calculation pipeline. 

Before each observing night (for OT), a set of calibration exposures is obtained. These include bias, darks, flat fields for order tracing, blaze function calculation, and pixel-to-pixel sensitivity correction. We also obtain flat fields through the iodine cell, which can be used to assess the long-term instrument stability.
For nightly observations, we define two basic modes: one for observations with the iodine cell inserted in the beam, and the other without. In the non-iodine mode, a ThAr spectrum is obtained just before and after the science exposure(s) to ensure a proper wavelength calibration. 

\begin{figure}
   \sidecaption
   \includegraphics[width=\columnwidth]{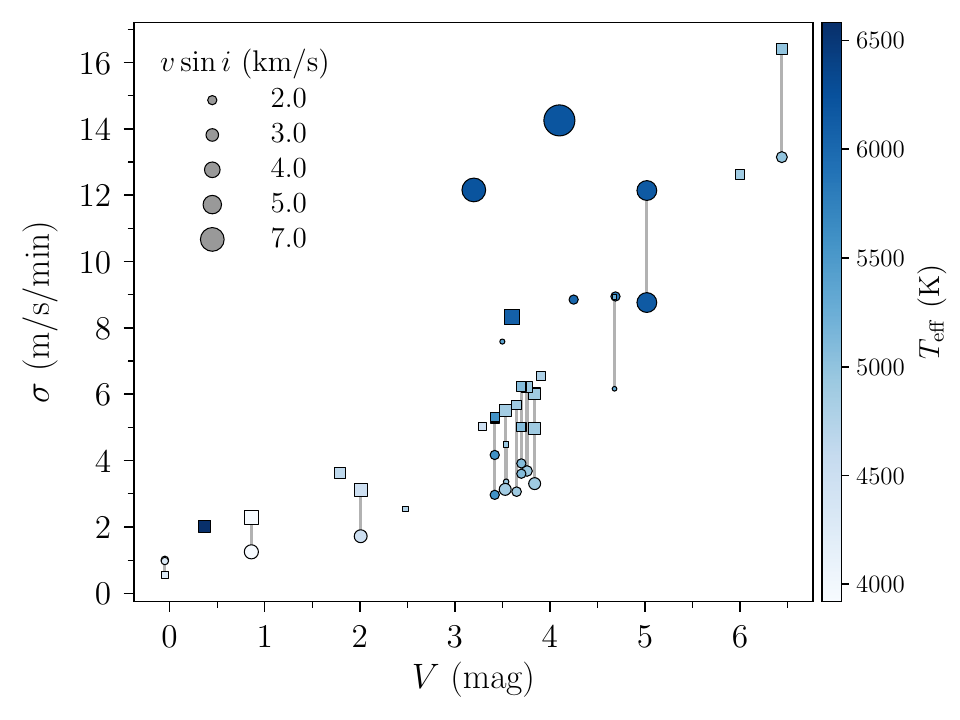}
      \caption{Relation between RV noise level (in m/s per minute) and stellar $V$-band magnitude for a sample of stars with time-series data from SONG OT. To rescale the measured per-exposure noise levels to per-minute values, we assume a scaling of noise level as $t_{\rm exp}^{-1/2}$. The circles indicate measurements taken with the current QHY detector, and squares indicate measurements taken with the ANDOR detector. Stars with noise values from both detectors are connected with a line. The marker colour indicates \teff, and the size indicates \vsini (see the legends).}
         \label{fig:noise_scale}
   \end{figure}

All spectra from Delingha and OT obtained before January 2024 used an Andor detector (\tref{tab:sites}) and the same extraction code (internally referred to as \code{songwriter}) developed by Jens Jessen-Hansen using \code{C++} modules from \citet{Ritter2014}.  
Following the change of the detector at OT in January 2024,
the \code{PyReduce} package is used \citep{Pyreduce2002, Pyreduce2020} for the spectral extraction at OT and MtK. A Python-based top-layer for \code{PyReduce}, called \code{songpipe}\footnote{\url{https://github.com/tronsgaard/songpipe}}, has been developed to handle naming conventions, instrument-specific tasks, and cases where calibration data are missing. All extracted spectra are inserted into SODA (see \sref{sec:access}).    

\begin{figure*}[!htbp]
   \sidecaption
   \includegraphics[width=\textwidth]{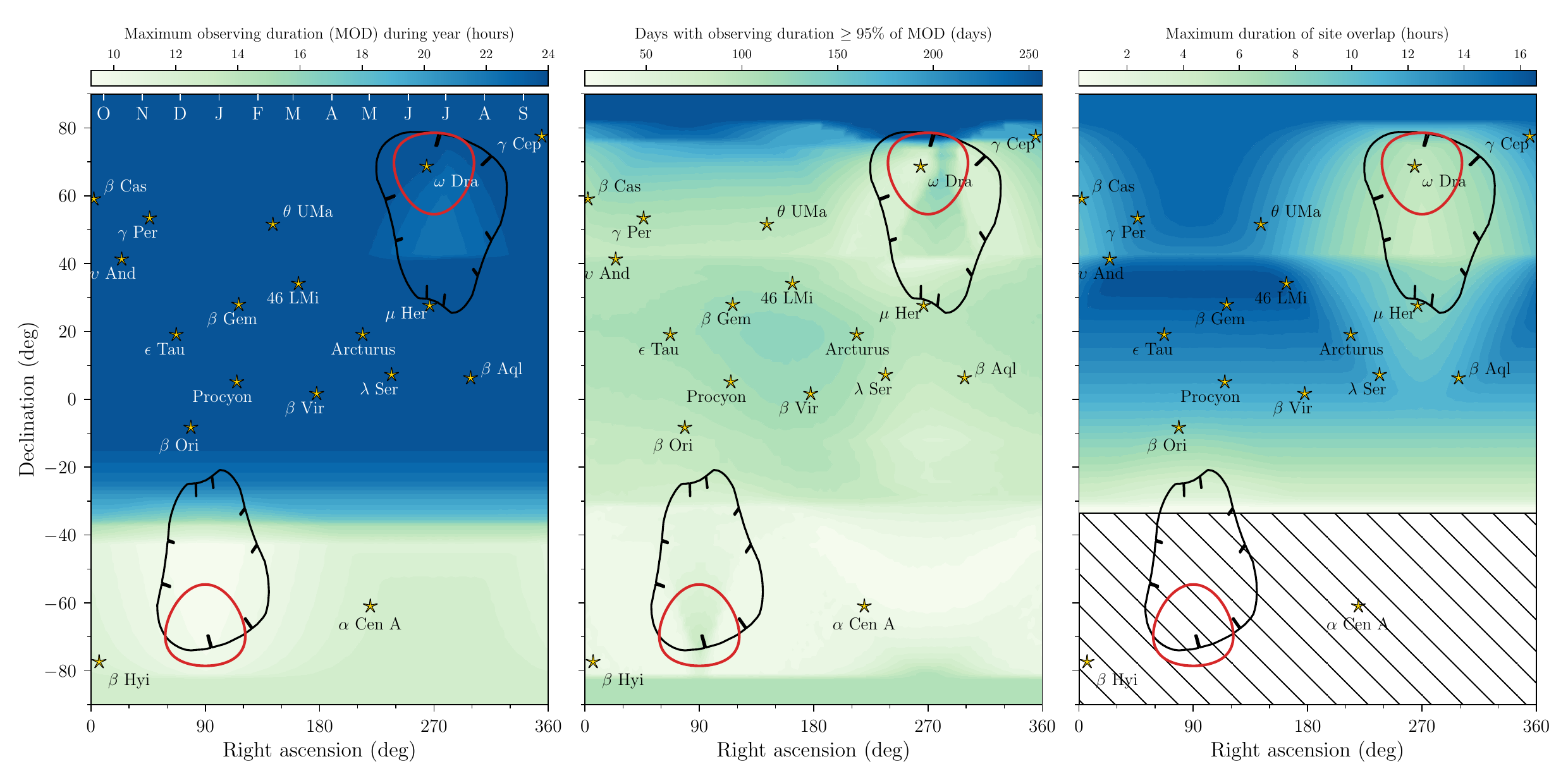}
      \caption{SONG multi-site observing coverage. \textit{Left}: Maximum possible daily observing duration (in hours; see the colour scale) across the year as a function of RA and Dec for stars observed with the full four-site (OT, Lenghu, MtK, and APO) SONG network. The ticks at the top of the panel indicate the time (in months by their first letter; starting with October (O)) during the year when a star with a given RA is highest in the sky in the middle of the night. \textit{Middle}: Number of days during a year when a star with a given RA and Dec is observable for a duration $\geq95\%$ of the maximum daily observing duration (left panel). \textit{Right}: Maximum possible daily overlap (in hours) between sites across the year as a function of stellar RA and Dec. The lower hashed region (at Dec $\lesssim-35^{\circ}$) indicates the region accessible only to the MtK site. In all panels, several specific stars are indicated, along with the outline of the PLATO LOP fields (LOPN1 and LOPS2) in black and the TESS continuous-viewing zones in red.}
         \label{fig:coverage}
   \end{figure*}
When a spectrum has been extracted, \code{Pyodine}\footnote{\url{https://gitlab.com/Heeren/pyodine}} \citep{Pyodine2023} is used to compute RVs from the iodine-cell data. \code{Pyodine} has setups for each site and detector combination and handles all required tasks, such as generating stellar templates and combining individual RV measurements into a final time-series product. 
For MtK, the data product consists of two extracted spectra per exposure (one from each telescope). For the subsequent \code{Pyodine} velocity measurements, these two are handled independently. 

\subsection{Data products}\label{sec:data}

\subsubsection{Spectra}\label{sec:spectra}

As outlined in \sref{sec:extract2}, the project has employed two different extraction codes (\code{songwriter} and \code{songpipe}), resulting in small differences in the data product formats.   

For \code{songwriter}, a total of five layers are returned. These include simple sum extraction, optimum-weight extraction, the blaze function, and wavelength solutions based on the ThAr spectrum closest in time to the stellar exposure, both before and after the science exposure.
\code{songpipe} returns five \code{FITS} records with names spec, sig, wave, cont, and columns to denote the extracted spectrum, uncertainty, wavelength, {\'e}chelle blaze, and the start and end columns for each extracted {\'e}chelle order. 

Furthermore, the \code{FITS} headers of the extracted spectra have evolved over time, whereby, for instance, keyword names may have changed, and there will be header differences depending on whether the spectrograph illumination is slit- or fibre-based.

\subsubsection{Radial velocity time series}\label{sec:timeseries}
Radial velocities are not currently a standard automatic output from SONG, but extracted RVs are available upon request for iodine-based observations. 

As a new initiative, RV time series (and other derived products) will become available on SODA for selected targets as so-called high-level science products (HLSPs). Going forward, we will provide all SONG data products used in published works as HLSPs, and we will gradually add analysis-ready RV time series, particularly for asteroseismic stars.   
\begin{figure*}
   \sidecaption
   \includegraphics[width=\textwidth]{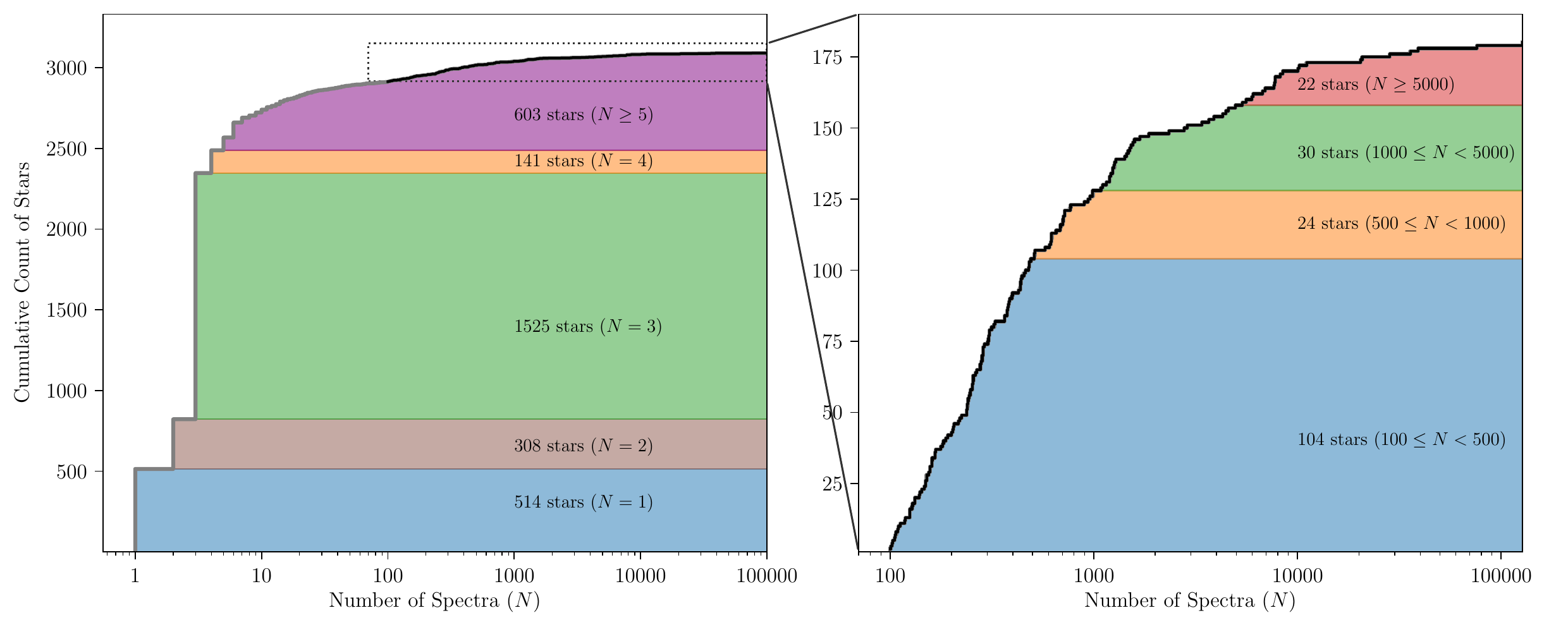}
      \caption{Cumulative distribution of stars against the number of spectra available in SODA. \textit{Left}: Full overview of all stars. The different colours highlight the different large blocks in the number of available spectra. \textit{Right}: Zoom-in on the top part of the CDF (left) for stars with at least 100 spectra.}
         \label{fig:cdf}
   \end{figure*}

\section{SONG observing capabilities and strategy}\label{sec:obs}

The characteristic noise level in SONG RV observations is central to evaluating which stars can be analysed for a specific scientific purpose.
In \fref{fig:noise_scale} we show the relation between the RV noise level (root-mean-square scatter in m/s per minute) and $V$-band magnitude for a sample of stars with time-series data from SONG OT.
As expected from the logarithmic relation between stellar flux and magnitude, the noise per unit time increases exponentially with increasing magnitude. In addition, the impact decreases with increasing \vsini. For example, the noise level of stars near $V{\sim}3.5$ is increased by a factor of ${\gtrsim4}$ when \vsini is increased from ${\sim}3$ to ${\sim}7\,$km/s. The improvement in the noise level across magnitudes is nearly constant at a factor of ${\sim}1.8$ when changing from the ANDOR to the QHY detector.  

We note that \fref{fig:noise_scale} is only representative of the SONG OT site. While the levels at the Lenghu site are still unknown, it is expected that APO, when fully operational, will reach uncertainty levels similar to that of OT. The MtK site has significantly higher uncertainties than OT (due to low fibre throughput); and therefore, when MtK observations are used for asteroseismology, they are only suitable for the very brightest stars ($V\lesssim3.7$; e.g. \object{$\alpha$ Cen~A}, \object{Procyon}, $\mu$~Her, and $\beta$ Aql; see \sref{sec:asteroseis}). 

In \fref{fig:coverage} we provide an overview of the observing coverage enabled by our future four-site network under the conditions of a minimum stellar altitude of $20^{\circ}$ and a Sun below $-6^{\circ}$ altitude. First and foremost, fewer gaps in the RV time series from the diurnal cycle are ensured, which is essential for reducing aliasing in the frequency analysis \citep{Arentoft2014}. Additionally, overlapping observations between sites reduce the risk of gaps due to bad weather and reduce RV noise from instrument- and site-specific systematics.

 \fref{fig:coverage} shows that for most stars with a declination (Dec) $\gtrsim35^{\circ}$, a full 24 h coverage is achievable over some period of the year (typically, for $\gtrsim100$ days), and all the time for (circumpolar) stars with Dec $\gtrsim80^{\circ}$. The pattern in the contours, which largely coincides with the northern PLATO candidate long-duration observation phase \citep[LOPN1;][]{Nascimbeni2022,Rauer2025} field, is caused by the missing (Pacific Ocean) coverage between the Lenghu and APO sites. We provide similar overviews in \aref{app:net} for two- and three-site network configurations.

\subsection*{Observing strategy}

The SONG scientific coordinators handle the overall prioritisation of targets for observation in close collaboration with the SONG working group (WG) chairs. The organisation of work on SONG data in terms of working groups, each associated with a specific scientific focus area (asteroseismology, exoplanets, binary stars, spectroscopic variability, or the Sun), was introduced in 2023 to replace a more standard approach with regular proposal calls and awarding of time to individual principal investigators. In the current setup, each working group can prioritise targets most relevant to their specific scientific focus, and the scientific coordinators then define the final balanced prioritisation of targets across WGs. Because of SONG's unique ability as a ground-based facility to obtain high-cadence observations over long baselines, asteroseismology is the primary driver of the network, and asteroseismic campaigns generally receive the highest priority in observational scheduling. Other types of observations are typically included as filler targets (see \sref{sec:wgs} for details on joining the SONG consortium and proposing new observations).

Based on the scientific prioritisation, the scheduling of observations during a given night is fully automatic and effectively handled by the \code{Conductor} software \citep{Andersen2019b}. Currently, the \code{Conductor} handles target scheduling at the OT, MtK, and APO sites and will be implemented at the Lenghu site when it is fully operational.

\section{SONG data overview (2012-2025)}\label{sec:overview}

\begin{figure*}
   \sidecaption
   \includegraphics[width=\textwidth]{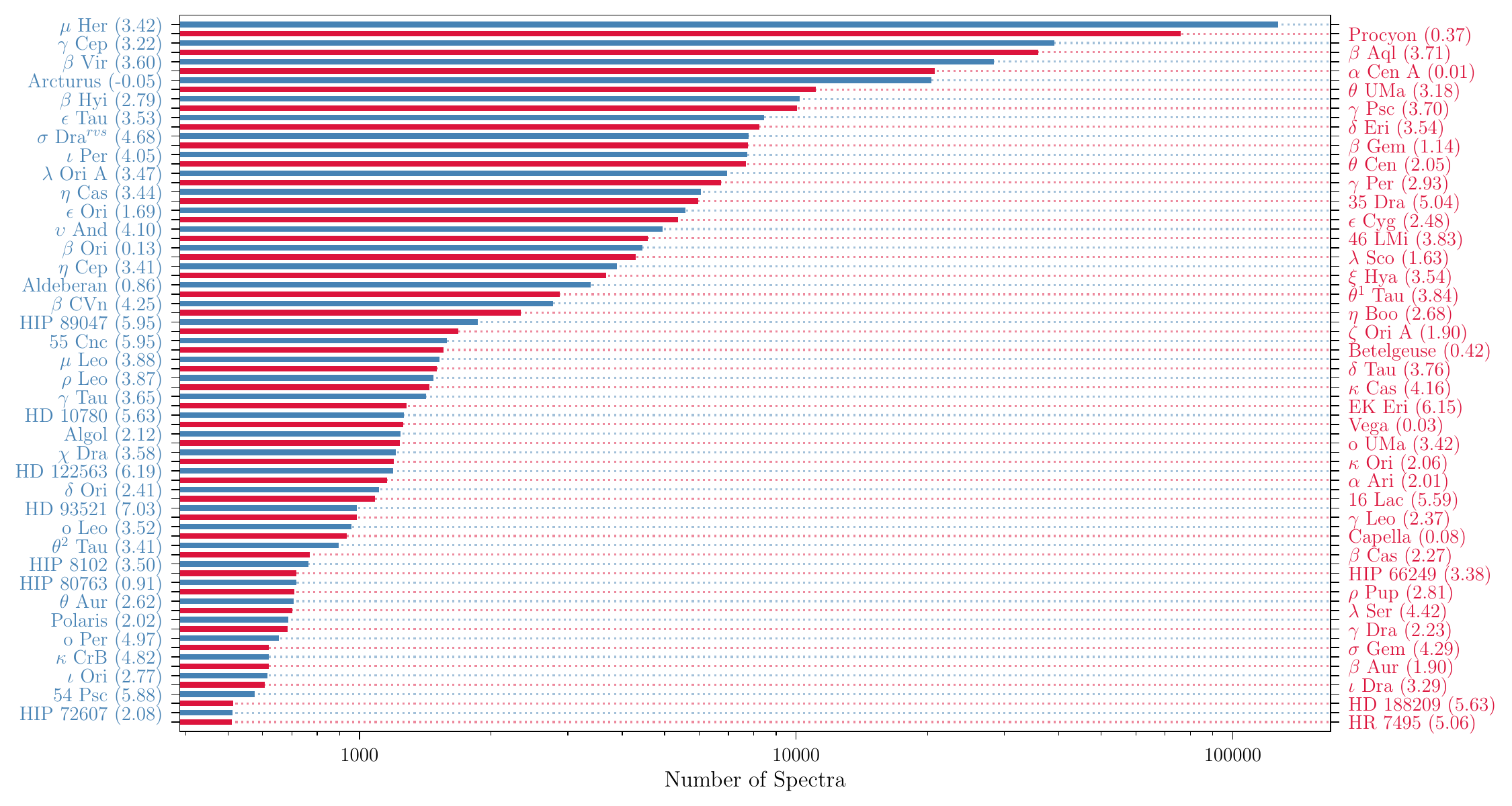}
      \caption{Bar chart of all stars observed by SONG with at least 500 spectra available in SODA as of the end of 2025. The number in parentheses after the star name indicates the visual stellar magnitude. The stars are alternately labelled on the left and right axes, with matching bar and label colours. The rvs for $\sigma$ Dra denotes its use as an RV standard star.}
         \label{fig:most_obs}
   \end{figure*}

\begin{figure}
   \sidecaption
   \includegraphics[width=\columnwidth]{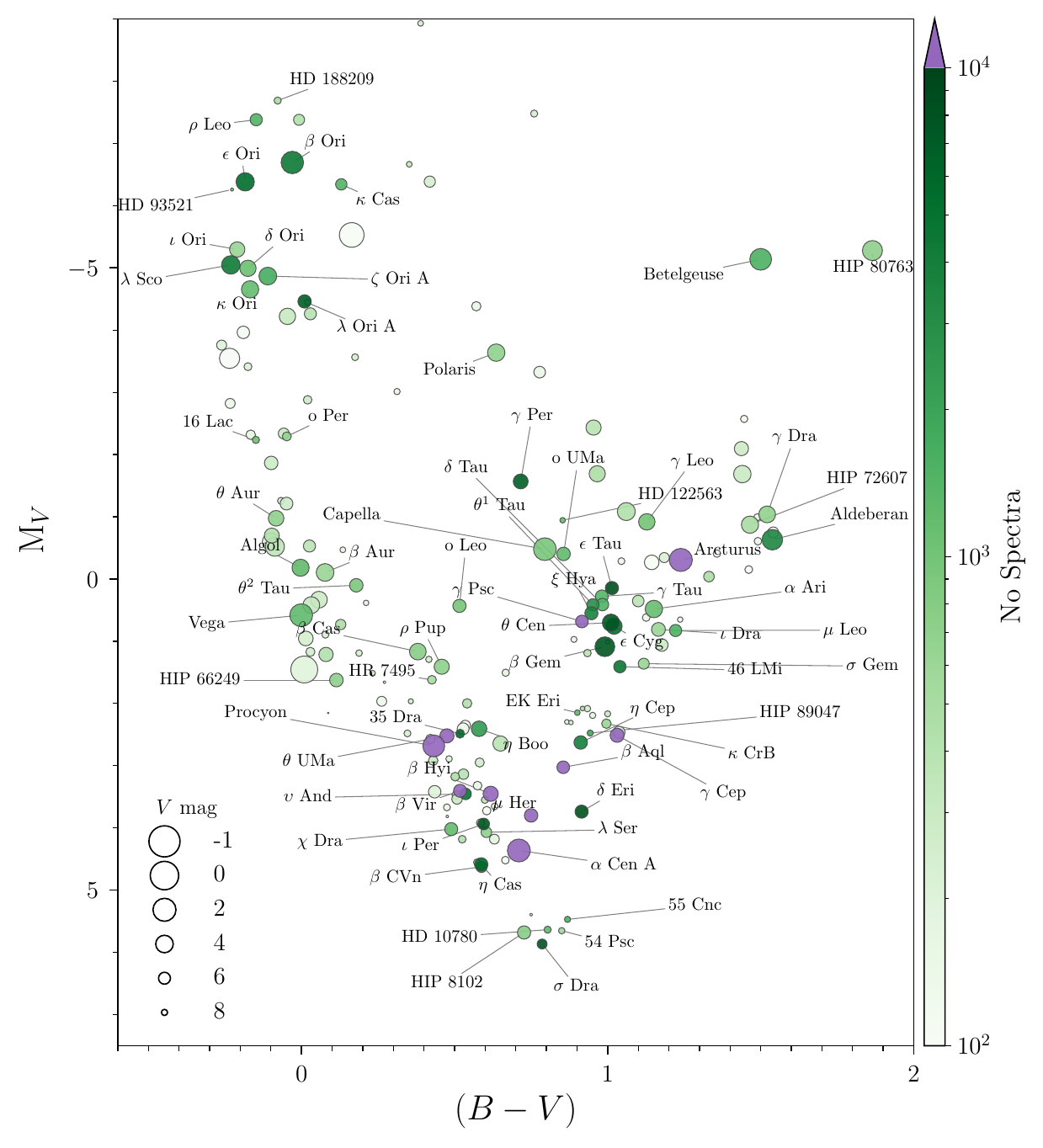}
      \caption{HR diagram showing the distribution of stars with ${\geq}100$ SONG spectra, and annotated for stars with ${\geq}500$ SONG spectra (see also \fref{fig:most_obs}). The stars are coloured according to the number of available spectra, and the size reflects the $V$-band magnitude (see the insert scale).}
         \label{fig:HRall}
   \end{figure}

\begin{figure}
   \sidecaption
   \includegraphics[width=\columnwidth]{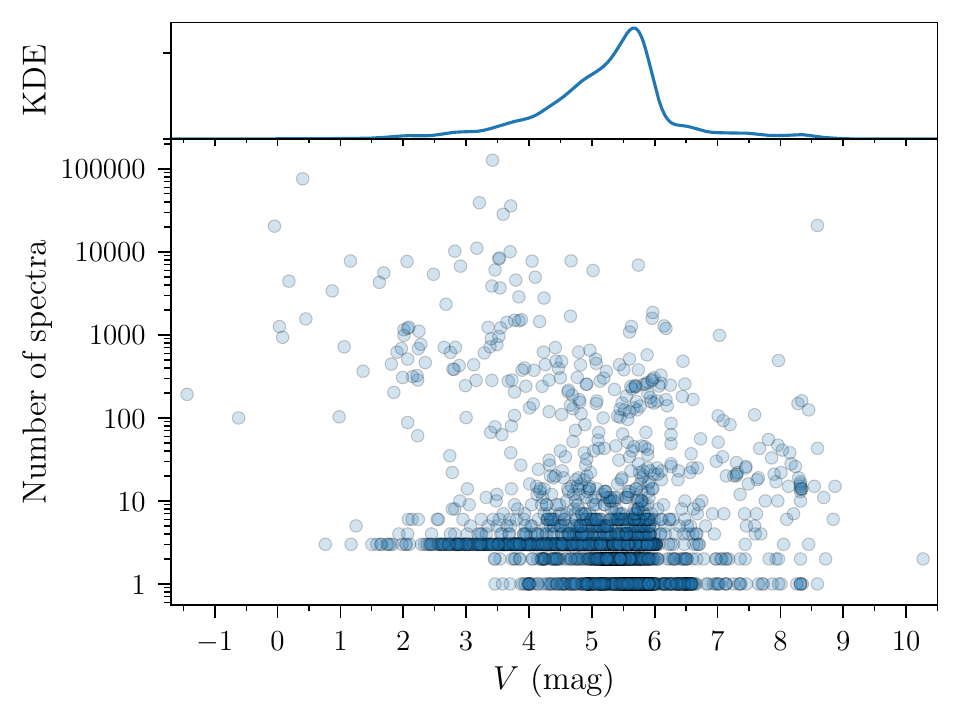}
      \caption{Correspondence between the visual ($V$) magnitude of stars observed by SONG and the number of spectra available in SODA. The top panel provides a kernel density estimate (KDE) of the stellar $V$-band magnitudes.}
         \label{fig:magdist}
   \end{figure}

Counting observations until the end of 2025, SODA contains a total of 583,\,005 spectra in raw and calibrated formats, divided across 3091 different stars. An additional 81,\,278 low-cadence solar spectra are also available.
Figure \ref{fig:cdf} provides an overview of the cumulative distribution of stars by the number of spectra available in SODA. As shown in the left panel, the majority of stars observed by SONG have between 1 and 5 spectra, which were mainly obtained for stellar atmospheric characterisation. The right panel shows that 180 stars have more than 100 spectra available. These are dominated by stars that are part of long-term monitoring programs for binary or exoplanet characterisation. At the tail of the distribution, 52 stars were observed more than 1000 times, and 22 of these have over 5000 observations. These targets represent the cohort of stars followed for asteroseismic characterisation (including the primary RV standard stars, \eg \object{$\sigma$ Dra}). Figure \ref{fig:most_obs} provides an overview of the stars observed by SONG with at least 500 spectra available, and \fref{fig:HRall} shows the distribution of these stars in an HR diagram. \fref{fig:magdist} shows the correspondence between visual magnitude and the number of available spectra; we see that the lion's share of stars observed by SONG are brighter than $V=6$. For asteroseismic analyses, main-sequence and sub-giant stars are generally brighter than $V=4$, while red giant branch stars can be analysed up to $V\approx6.5$ \citep{Stello2017,Malla2020}.   

The majority ($\gtrsim88\%$) of currently available observations in SODA are from SONG OT, which has been operational for more than a decade (see \tref{tab:sites}). Starting in April 2021, observations from MtK contributed another ${\sim}10\%$, while the initial Chinese node at Delingha, operated manually from November 2016 to December 2019, contributed the other ${\sim}2\%$ of observations.
In terms of observing modes, ${>}75\%$ of observations have been obtained via the iodine cell, while the remaining observations have been obtained using ThAr. The ThAr observations are mainly obtained for binaries for which iodine-mode spectrum analysis is challenging, or for single-shot observations for spectroscopic characterisation.  

Many stars have observations from multiple sites, currently restricted to OT, MtK, APO (from late 2025), and Delingha. In terms of using this core capacity of the network, the main targets are those of interest to asteroseismology (see \sref{sec:asteroseis}), specifically, \object{$\mu$ Her} \citep{Grundahl2017}, \object{$\gamma$ Cep} \citep{Knudstrup2023}, \object{$\beta$ Aql} \citep{Kjeldsen2025}, and \object{Arcturus} (\object{$\alpha$ Boo}; \citealt{Tarrant2007}; Pall{\'e} et al., in prep.).   

\section{SONG science}\label{sec:science}
\begin{figure}
\centering
   \includegraphics[width=\columnwidth]{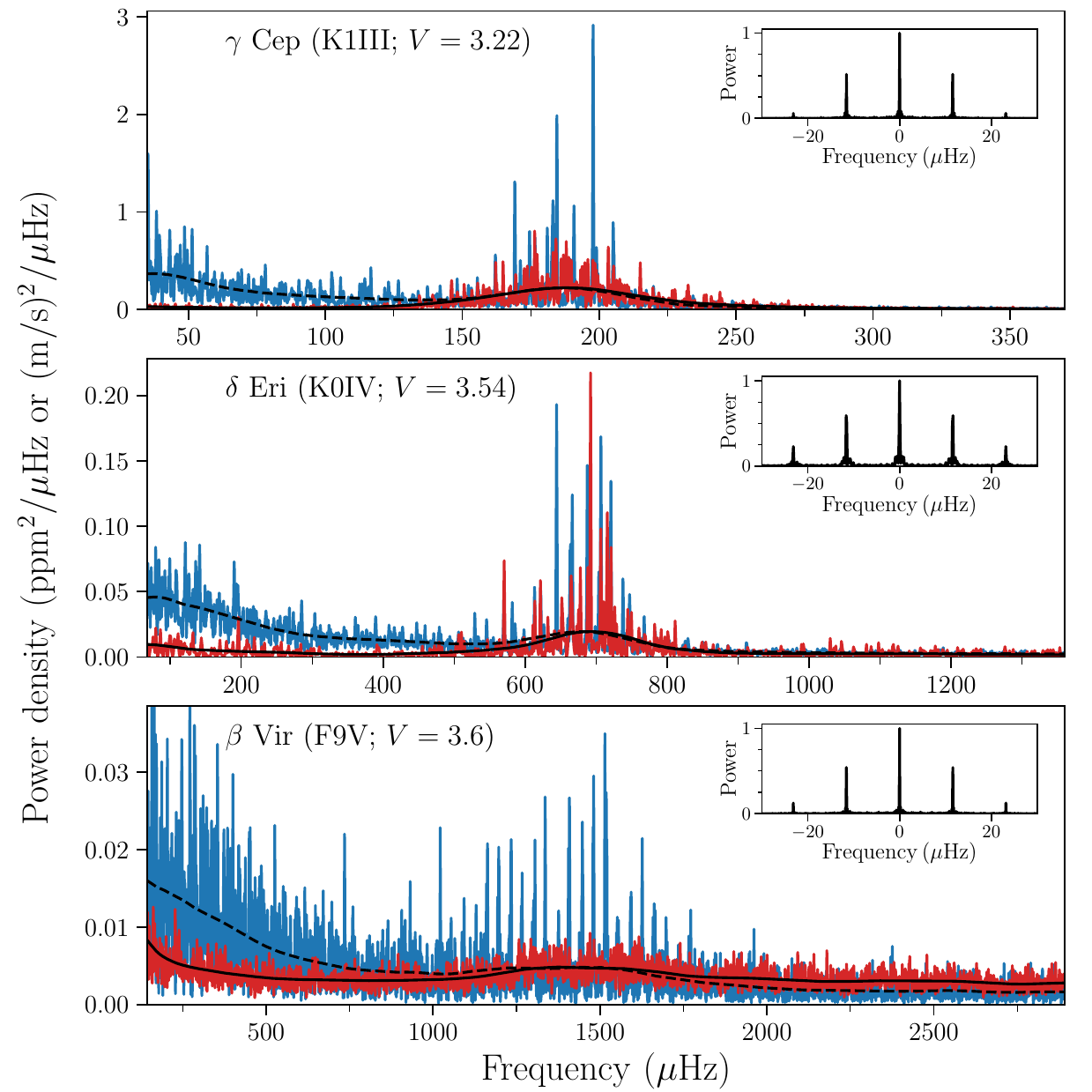}
      \caption{Examples of power density spectra for stars with detectable asteroseismic excess from SONG RV observations: $\gamma$ Cep (\textit{top}), $\delta$ Eri (\textit{middle}), and  $\beta$ Vir (\textit{bottom}). The spectra in red are based on available SONG-OT observations, and the corresponding spectrum based on TESS 20 s cadence photometry is shown in blue. To highlight the different impacts of low-frequency granulation noise, the TESS spectra have been normalised to match the power density level from SONG near \numax, based on the heavily smoothed spectra (using an Epanechnikov filter with a width of $4\dnu$) given by the full (SONG) and dashed (TESS) black lines. The inserts show the spectral window from the SONG observations for each star.}   
         \label{fig:ps_ex}
   \end{figure}
\subsection{Asteroseismology of solar-like oscillators}\label{sec:asteroseis}
The ability to conduct long-term monitoring of individual stars makes SONG a unique facility for ground-based asteroseismic studies because the long baseline provides a high-frequency resolution and the option to track stellar magnetic activity cycles through their impact on asteroseismic parameters \citep{Garcia2010,Kiefer2017,Santos2018,Santos2019}. Other important benefits of observations in velocity as opposed to photometry are (1) the well-known reduction in the impact of low-frequency granulation noise \citep[\eg][]{Harvey1988,Kjeldsen2011,Jackiewicz2021}, which is evident from \fref{fig:ps_ex}. This enables the detection especially of lower-frequency modes, which are particularly important in asteroseismic modelling as they are less affected by improper modelling of the outer surface layers \citep{Ball2017}. (2) The generally higher visibility of modes facilitates the detection especially of octupole $l=3$ modes \citep{Schou2018}.    

Given the dependence of RV noise levels on stellar brightness, as shown in \fref{fig:noise_scale}, and the corresponding integration times required to obtain high S/N RV measurements, asteroseismology with SONG is generally limited to subgiants or giants for stars with $V{\lesssim}5$. For fainter stars, the required integration time typically attenuates the oscillation signal too severely or moves the Nyquist frequency below asteroseismic mode excess. The requirement for observing bright stars with SONG does, however, provide a natural complement to some of the photometric missions suitable for asteroseismology, such as \textit{Kepler} \citep{Gilliland2010} and TESS, as these often saturate heavily for the brightest stars, requiring specialised treatments, such as halo or smear photometry \citep{White2017,Pope2019b,Pope2019, Rudrasingam2026}. 

With these observational advantages in mind, the SONG network sites contribute to asteroseismic studies as follows. The main site for asteroseismic programs is OT. MtK observations primarily focus on the few bright southern solar-like oscillators such as $\alpha$ Cen, Procyon ($\alpha$ CMi), and $\beta$ Hyi. Observations have also been obtained at the original Delingha site in China, but mainly for the stars $\gamma$ Cep, $\mu$ Her, and $\beta$ Aql. While most observations to date have been conducted in single-site mode, the quality of the RV time series will improve as more sites are added to the network. 

\citet{Grundahl2017} provided the first asteroseismic results from SONG for the subgiant star \object{$\mu$ Her}, which was first studied seismically by \citet{Bonanno2008} based on seven observing nights. The multi-star system has been observed since then and now comprises more than 12 seasons of data (Kjeldsen et al. in prep.). \citet{Grundahl2017} clearly indicated the potential for detailed asteroseismic analyses based on SONG data, even in single-site mode, and modelling of the asteroseismic parameters has since been pursued by several groups \citep[see][]{Li2019,Gupta2025}. The value of $\mu$ Her as an asteroseismic benchmark has recently been substantiated by \citet{Marcussen2026}, who provided independent dynamical mass determinations at ${\sim}1\%$ precision from detailed binary modelling that incorporated RVs (including SONG) and astrometric visual binary measurements. 

Following the initial analysis of $\mu$ Her, several asteroseismic investigations of individual stars have been conducted, including the study of the red giants \object{46 LMi} 
\citep{Frandsen2018}, \object{Aldebaran} ($\alpha$ Tau; \citealt{Farr2018,Beck2020}), and the metal-poor giant \object{HD 122563} \citep{Creevey2019}, and the planet-hosting giants $\epsilon$ Tau in the Hyades open cluster \citep{Arentoft2019} and \object{$\gamma$ Cep} \citep[][see also \fref{fig:ps_ex}]{Knudstrup2023}.

Asteroseismic masses were derived based on SONG data for eight evolved planet-hosting stars\footnote{$\gamma$ Cep, $\epsilon$ Tau, $\beta$ Gem (Pollux; potentially not a planet-host according to \citealt{Auri2021}), 18 Del, HD 5608, $\kappa$ CrB, 6 Lyn, and HD 210702.} by \citet{Stello2017} and later followed up and augmented with four\footnote{24 Sex, HD 167042, HD 192699, and HD 200964.} additional stars by \citet{Malla2020}. The so-called retired A-star controversy \citep{Johnson2006,Lloyd2011,Lloyd2013} concerns the potential overestimation of red giant masses derived from spectroscopy and, hence, the proper identification of A-star descendants for exoplanet population studies. \citet{Stello2017} and \citet{Malla2020} both sought to address this (at the time) controversy, finding that spectroscopic masses are indeed overestimated compared to asteroseismology, but in a mass-dependent manner affecting stars with masses above ${\sim}1.6\,\msol$ \citep[see also][]{North2017}.

Recently, results for \object{$\beta$ Aql}, another key asteroseismic subgiant for the SONG project, were presented by \citet{Kjeldsen2025} based on two seasons of data. In addition to seismic modelling based on individual mode frequencies, this analysis also highlighted a very promising avenue for the future use of SONG. Using contemporaneous TESS photometric observations, the authors calculated the relation between RV and photometry for oscillation amplitudes and phases across a range of different modes (see also \citet{Arentoft2019} for a non-contemporaneous amplitude comparison for the Hyades giant $\epsilon$ Tau). 
Such relations inform our understanding of oscillation-mode physics and of mode excitation and damping \citep{Houdek2010,Beck2020,Zhou2021}. Furthermore, when these amplitude and phase relations are robustly established, photometric observations can be used to correct more expensive and sparse RV observations for the impact of oscillations \citep[similar to the treatment of, for instance, spots by][]{Aigrain2012}. This would be very valuable to exoplanet searches, where the stellar intrinsic signals (`noise' to the exoplanet community) currently set the boundary for the available precision \citep{Dumusque2015,Fischer2016,McWilliam2026}.

It is a focus of current and future SONG observations for asteroseismology to exploit this synergy. With TESS, we currently have a good overview of the seismic targets of interest \citep{Lund2025,Rudrasingam2026}, and this will also be the case for bright stars observed by PLATO \citep[][]{Rauer2025,Eschen2024,Nascimbeni2025,Nascimbeni2026,Panetier2026}.
Furthermore, the comparison between RV and photometric data might facilitate new studies investigating poorly understood properties of oscillation modes, such as mode profile asymmetries \citep{Duvall1993,Nigam1998,Benomar2018,Philidet2020} and visibilities \citep{Salabert2011,Lund2014,Lund2017,Schou2018}.  

We note that asteroseismic detections have been made in several solar-like oscillators that are amongst the most observed by SONG (see \fref{fig:most_obs}), including \object{$\beta$ Vir}, \object{$\alpha$ Cen A}, \object{$\delta$ Eri} (\fref{fig:ps_ex}), \object{$\beta$ Hyi}, \object{Arcturus} (\object{$\alpha$ Boo}), Procyon ($\alpha$ CMi), $\theta$ UMa, 35 Dra, and many more. Many of these are still awaiting publication, but the availability of accompanying TESS photometry (\fref{fig:ps_ex}) for a large number of targets \citep{Lund2025,Rudrasingam2026} opens promising prospects for the joint RV–photometric analyses described above. 

\subsection{Binaries}
\begin{figure*}[htbp!]
   \sidecaption
   \includegraphics[width=\textwidth]{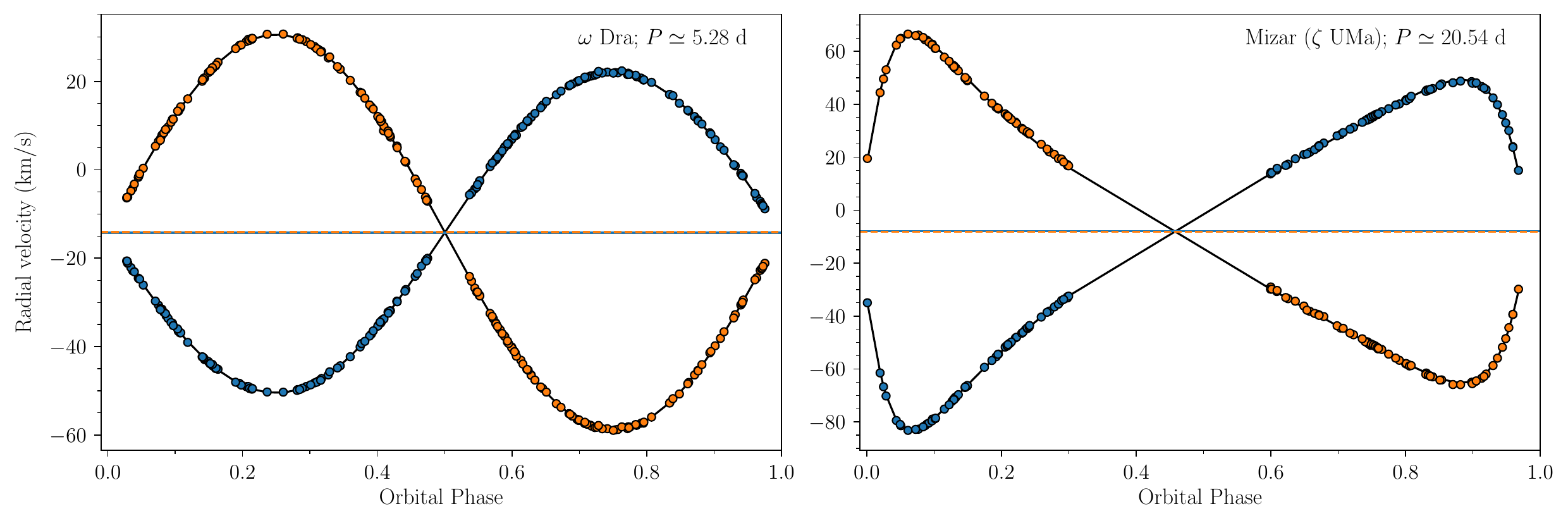}
      \caption{Examples of SONG filler observations of the binaries $\omega$ Dra (\textit{left}) and Mizar ($\zeta$ UMa; \textit{right}), with model fits from Wang et al. (in prep.). The RVs belonging to two components are shown with blue and orange markers. Correspondingly, the horizontal full blue (dashed orange) line indicates the individual component system velocities, accounting for potential differences in gravitational redshift.}
         \label{fig:binaries}
   \end{figure*}
Monitoring of binary star systems is ideally suited to SONG, owing to the long-term coverage that benefits asteroseismology and the ease of adding new targets for filler or backup observations. While SONG is primarily focused on high-cadence coverage for asteroseismology, time is available for other targets at the beginning and ends of most nights. Small gaps are also inserted in the primary target coverage when deemed inconsequential for the asteroseismic analysis, and poor-weather conditions occasionally provide additional opportunities.

A large sample of bright stars known to be spectroscopic binaries was targeted for filler observations early in the network operations. Many now have semi-regular observations spread across the past decade. As examples, $\omega$ Dra and $\zeta$ UMa (Mizar), both part of this early program, are shown in \fref{fig:binaries}. While observations from this long-term monitoring may not lead to the discovery of new systems, orbital parameters can often be significantly improved. The results in \fref{fig:binaries} are part of a catalogue analysis underway by Wang et al. (in prep.). 

Binary systems benefit particularly from such observations when one or both components oscillate, and both stars are spectroscopically resolved. If information about the system inclination can be obtained, individual dynamical masses can be measured, providing important anchoring for masses obtained from asteroseismology \citep{Serenelli2021}. Such systems have so far mainly comprised rare eclipsing red giant binaries \citep{Gaulme2016,Brogaard2018,Brogaard2021,Brogaard2022,Benbakoura2021,Thomsen2022,Thomsen2025}, in which the inclination can be determined from eclipse modelling. However, astrometric observations can provide equivalent information via an on-sky trace of the orbit, enabling the use of (the currently sparse sample of) wider, less-evolved systems for benchmarking asteroseismology \citep{Ball2022,Beck2026}. 
Notable non-eclipsing systems that have benefited greatly from long-term SONG observations include, first, $\chi$~Dra, whose main-sequence primary was recently found to exhibit solar-like oscillations in TESS data \citep{Lund2025}. A full comparison of asteroseismic and dynamical masses will be presented by \citet{Rudrasingam2026sub}. Second, the subgiant $\mu$ Her Aa \citep{Grundahl2007}, where 12 years of SONG spectroscopy, combined with literature RV and interferometric data, has enabled a full solution of the hierarchical quadruple system \citep{Marcussen2026}, providing yet another benchmark for asteroseismic masses.   

Beyond these two systems, \citet{Lund2025} recently provided the TESS Luminaries Sample of bright ($V\leq6$) solar-like asteroseismic targets detected in TESS photometry, several of which are known double-lined spectroscopic binaries with available visual-orbit astrometric measurements (see their Table~5). Most of these systems are now being actively followed by SONG and promise to substantially extend the current sample of MS and SG benchmarks for asteroseismology.

Beyond asteroseismic applications, SONG contributes to a broad range of binary-star programmes, including single-component spectroscopic analyses such as that of $\alpha$ Dra \citep{Hey2022}, investigation of the binary properties of Galactic red supergiant stars \citep{Patrick2020,Patrick2025}, follow-up of massive OB-type binaries as part of the large IACOB survey \citep{Simondiaz2014,SimonDiaz2024}, and southern hemisphere monitoring of long-period evolved binaries using SONG-MtK, complementing corresponding HERMES at Mercator \citep{Raskin2014} observations in the north \citep{Escorza2020,Escorza2023}. SONG data have also been used to assess binarity in hump-and-spike stars \citep{Saio2018}, a sample of ${\gtrsim}200$ A- and F-type stars showing Rossby modes and rotational modulation \citep{Henriksen2023b,Henriksen2023,Antoci2025}.

\subsection{Exoplanets}
SONG-OT data contributed to confirming the transiting hot Jupiters MASCARA-1b \citep{Talens+2017}, MASCARA-2b \citep{Talens+2018}, MASCARA-3b \citep{Hjorth2019}, and TOI-1431b/MASCARA-5b \citep{Addison2021}. The first three planets have been observed in time series spanning multiple hours just before, during, and after a planetary transit in front of their host stars, enabling measurements of the Rossiter–McLaughlin effect \citep{Rossiter1924, McLaughlin1924}. This allowed for measurements of the projected stellar obliquity. 

Shorter gaps ($<$1 hour), when asteroseismic targets are unobservable, can readily be filled with single-epoch measurements of exoplanet candidates, enabling RV measurements and planet mass determinations. This approach has also proven useful for testing the validity of candidate signals from other surveys of bright evolved stars \citep{Luque+2019,Heeren2021,Spaeth2025}. Using RV from OT, \citet{Wittenmyer2026} recently confirmed the presence of a 1.67$M_{\rm Jup}$ giant planet orbiting the low-luminosity giant HD\,126105. In a similar vein, low-cadence SONG RVs are useful for monitoring the orbits of known multiple-giant-planet systems to detect long-term dynamical interactions among the planets.  
We also make a point of sparsely monitoring most stars targeted for asteroseismology to ensure that we can track the long-term stability of their velocities and potentially detect new exoplanet companions.

\subsection{Stellar variability and other asteroseismic targets}

\subsubsection{Spectroscopic variability of massive stars}\label{sec:otype}

The first SONG observations of massive OB-type stars were obtained with OT in December 2014 and January 2015. As part of a pilot programme designed to assess the capabilities of the spectrograph for detecting and characterising line-profile variability in massive stars, we selected four bright (see \fref{fig:HRall}) well-studied targets in Orion: Rigel ($\beta$\,Ori, B8\,Ia), Saiph ($\kappa$\,Ori, B0.5\,Ia), Alnilam ($\epsilon$\,Ori, B0\,Ia), and Meissa ($\lambda$\,Ori\,A, O8\,III((f))).

This first campaign successfully demonstrated SONG's potential for investigating the diverse spectroscopic variability exhibited by O-type stars and B supergiants \citep[\eg][]{SimonDiaz2024}. In particular, the combination of high temporal resolution, long nightly coverage, and excellent instrumental stability makes SONG ideally suited for detecting low-amplitude pulsational variability and for resolving complex beating patterns expected from multi-period oscillations \citep[\eg][]{Fullerton1996, Pamyatnykh1999, Aerts2009, SimonDiaz2017, Godart2017, Burssens2020, Bowman2023}.

\begin{figure}
\centering
   \includegraphics[width=\columnwidth]{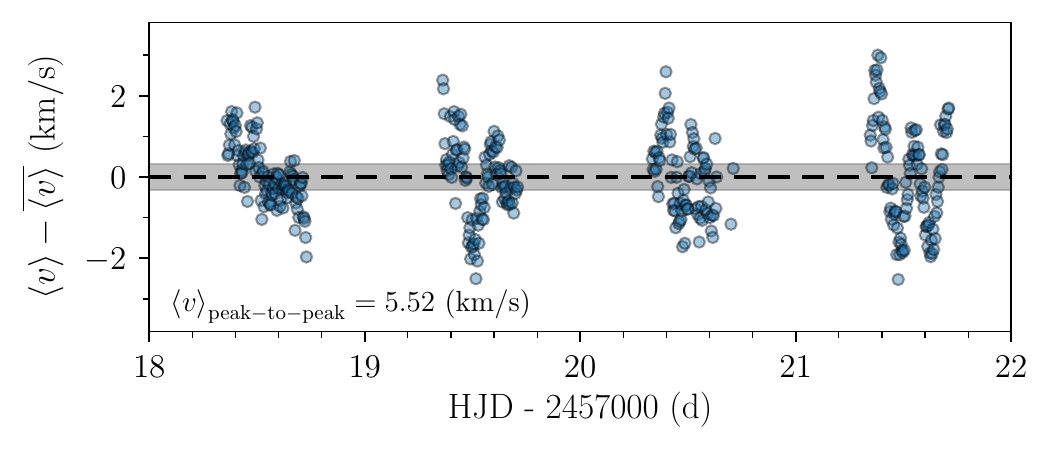}
      \caption{First detection of a clear variability pattern in the first moment ($\langle v \rangle$) of the He\,{\sc i}~5015 line profile of the O8 giant star HD\,36861 (Meissa, $\lambda$\,Ori\,A), relative to nightly median levels. The horizontal grey band highlights the robust standard deviation ($1.4826$ times the median absolute deviation) of the point-to-point scatter in $\langle v \rangle$ (at a value of $0.64$ km/s), and the value of the maximum peak-to-peak variation is indicated in the plot.}   
         \label{fig:lambdaOri}
   \end{figure}

Beyond confirming known variability patterns in Rigel and Alnilam \citep[][]{Moravveji2012, Prinja2004}, the campaign yielded the first detection of clear spectroscopic variability in the photospheric lines of the O8 giant $\lambda$\,Ori\,A (\fref{fig:lambdaOri}). In addition, it highlighted the need to extend the overall temporal baseline of the observations to characterise the variability of the different wind-sensitive diagnostic lines available within the SONG spectral window, namely H${\alpha}$, H${\beta}$, He\,{\sc i}~5875, and He\,{\sc ii}~4686 \citep[\eg][]{Prinja2004, Martins2015, SimonDiaz2018}.

SONG has thus emerged as a natural ground-based complement to TESS photometry for hundreds of Galactic OB stars \citep[\eg][]{Burssens2020}. The combination of time-resolved spectroscopy and space photometry opens new avenues for asteroseismology of massive stars, where disentangling pulsations from wind variability and rotational modulation is essential for constraining their internal structure.

\begin{figure*}
\sidecaption
   \includegraphics[width=0.7\textwidth]{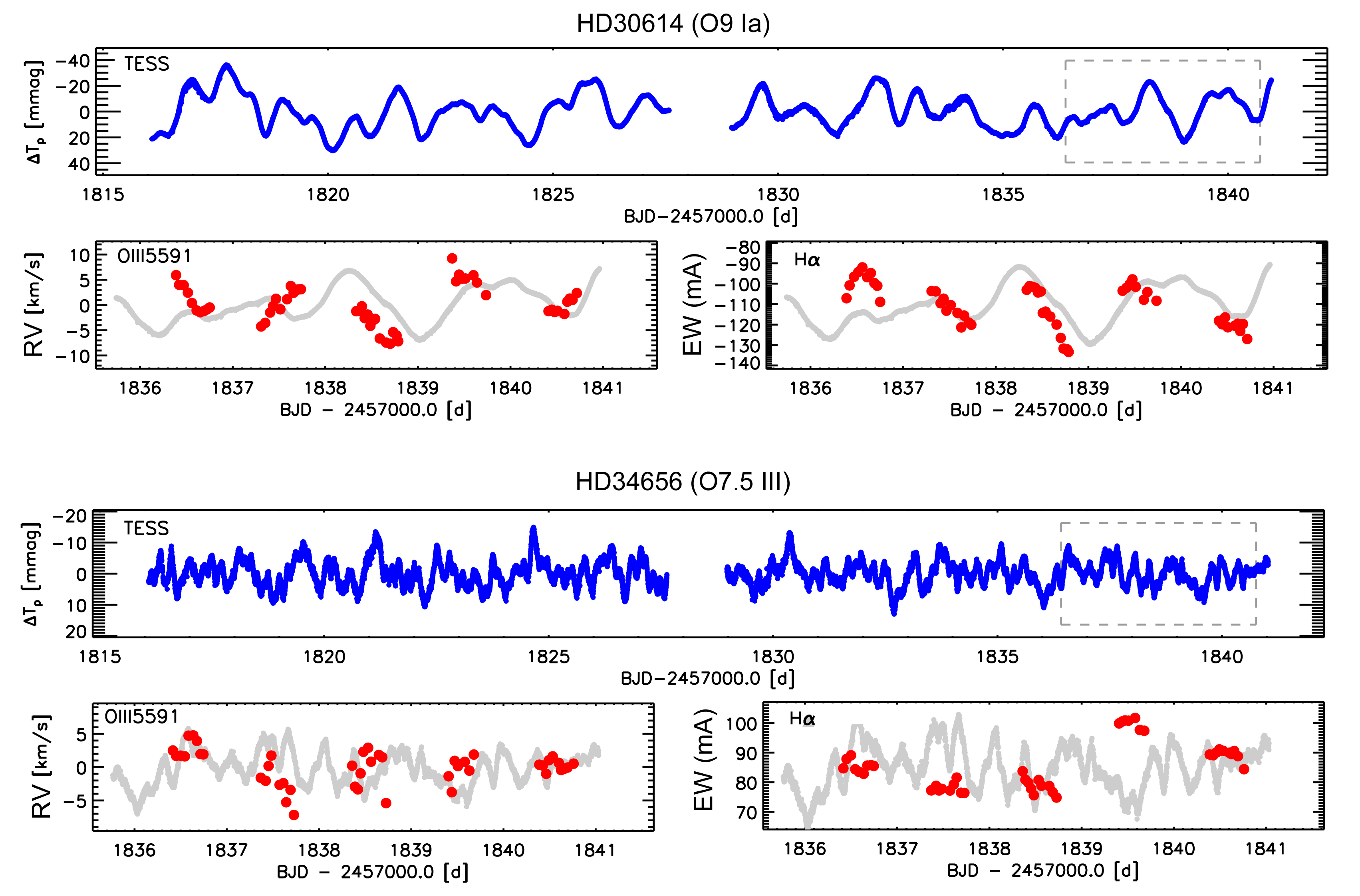}
      \caption{Contemporaneous observations obtained with TESS and SONG for the two O-type stars \object{HD 30614} ($\rho$\,Leo) and \object{HD 34656}. \textit{Top panels}: Photometric variability from the TESS mission over $\sim$30~days (in blue). The dates when SONG was gathering high-cadence full-night observations of the two stars are indicated by the dashed grey rectangle. \textit{Bottom panels}: Spectroscopic variability of each star (red dots), with RV variations detected in the O~{\sc iii}\,5591 line (\textit{left}) and equivalent width variations of the wind-diagnostic line H$_{\alpha}$ (\textit{right}). The TESS photometry is shown as the grey line.\vspace{2cm}}    
         \label{fig:OBstars-TESS}
   \end{figure*}

Over the years, SONG has conducted several observing programmes on massive OB-type stars, with allocations of approximately 150--250~h per semester. Specifically, SONG enabled the monitoring of ten O-type giants and supergiants and 12 B supergiants over timescales from days to months, with cadences ranging from minutes to hours. Parts of these observations were obtained contemporaneously with TESS (\fref{fig:OBstars-TESS}), while others could be combined with photometric data from the {\em Kepler} and K2 \citep{Howell2014} missions.

Illustrative examples of the types of studies enabled by the compiled SONG observations can be found in \citet{Aerts2017,Aerts2018} and in \citet{SimonDiaz2018}, focusing on the late O- and B-type supergiants HD~188209 (O9.5\,Iab), $\rho$~Leo (HD~91316, B1\,Iab), and $\kappa$~Cas (HD~2905, B1\,Ia). While these studies benefited exclusively from spectroscopic time series obtained with a single SONG node, we have in recent years initiated coordinated observing campaigns involving the three currently operational nodes (Caballero-Almagro et al., in prep.).

\subsubsection{Spectroscopic variability due to magnetic activity}\label{sec:mag}

Cool stars exhibit phenomena caused by surface magnetic fields, such as photospheric cool starspots and chromospheric plages. Signatures of this activity are visible in high-resolution spectral line profiles, varying on the rotational timescale (days to tens of days), with longer-term modulation from spot evolution (months) and activity cycles (years).
SONG enables continuous observations over a full stellar rotation, allowing the study of magnetic features across the stellar surface. Depending on the stellar rotation period, observations can also be carried out over several rotations within a single observing season, enabling investigation of the evolution of magnetic features from one rotation to the next. Over the years, SONG has observed several well-known active stars, most notably, $\sigma$~Gem (619 spectra), $\zeta$~And (375 spectra), $\sigma^{2}$~CrB (361 spectra), and IM~Peg (67 spectra).

Time series of SONG data are especially well suited to studying the location and evolution of temperature spots on cool stars using Doppler imaging. Doppler imaging maps stellar surface temperature or abundance from rotationally modulated distortions in high-resolution line profiles \citep[\eg][]{Vogt1987, Piskunov1990}, caused by inhomogeneous temperature distributions, such as starspots, or chemical abundance patterns as seen in chemically peculiar A- and B-type stars.

The first study of stellar magnetic activity using SONG data mapped the stellar surfaces of the two components of the RS\,CVn-type active binary $\sigma^{2}$\,CrB using Doppler imaging \citep{Xiang2020}, based on SONG data obtained over 11 nights in 2015. The resulting Doppler images showed dominant polar spots on the F9 and G0 star components of $\sigma^{2}$\,CrB, consistent with earlier studies. Based on the two independently reconstructed Doppler images, \citet{Xiang2020} also recovered solar-like surface differential rotation on the G0 binary component, with the equator rotating faster than the poles.

The second published Doppler imaging study using SONG data concentrated on single-lined RS\,CVn-type binary $\sigma$\,Gem \citep{Korhonen2021}. 
The data span 150 nights and eight consecutive stellar rotations in late 2015 and early 2016, yielding eight independent temperature maps that track surface features from one rotation to the next. These maps reveal predominantly high-latitude and polar spots above $45^{\circ}$, migrating from orbital phase 0.25 at the start of observations to phase 0.75 by the end.

SONG data can also be used to investigate chromospheric activity using lines formed in the chromosphere, such as H$\alpha$, H$\beta$, the $\mathrm{Na\,I}$ D1, the D2 doublet, and $\mathrm{He\,I}$ D3 lines. \cite{Cao2022} investigated chromospheric activity in the RS\,CVn-type binaries $\sigma$\,Gem and IM\,Peg. The authors confirmed that both stars are very active, although the IM\,Peg activity level is much higher on average. The chromospheric lines also exhibit rotational modulation and asymmetric line shapes, indicating the presence and evolution of chromospheric activity features.

Finally, the SONG data can also be used to study the interplay of oscillations and magnetic fields. \cite{Bonanno2019} observed the over-active G8 sub-giant EK\,Eri with SONG, using 312 spectra obtained over about eight consecutive nights in December 2017. The observations presented further evidence of acoustic oscillations on EK\,Eri and \cite{Bonanno2019} determined a large separation for the first time on this target. As this star is over-active for its spectral type, it is also a unique laboratory for studying the behaviour of acoustic pulsations in the presence of a large-scale magnetic field. 

\subsection{Solar-SONG}

Initial tests of SONG's capabilities for solar observations were conducted in 2012 at OT during a delay in the telescope installation \citep{Palle2013}. The light was fed to the spectrograph via a $400\,\mu$m optical fibre pointed at the Sun, and observations were obtained during a week in June 2012. The comparison of observations from SONG with those obtained contemporaneously with the BiSON Mark-I instrument \citep[][also located at Observatorio del Teide]{Chaplin1996} and the SoHO GOLF instrument \citep{Garcia2013} suggested that SONG performed very well, with high-frequency ($\rm 6000$--$\rm 8000\, \mu Hz$) noise levels lower by factors of 2.5 and 4.4, respectively \citep{Breton2022}. 
\begin{figure}
   \includegraphics[width=0.5\textwidth]{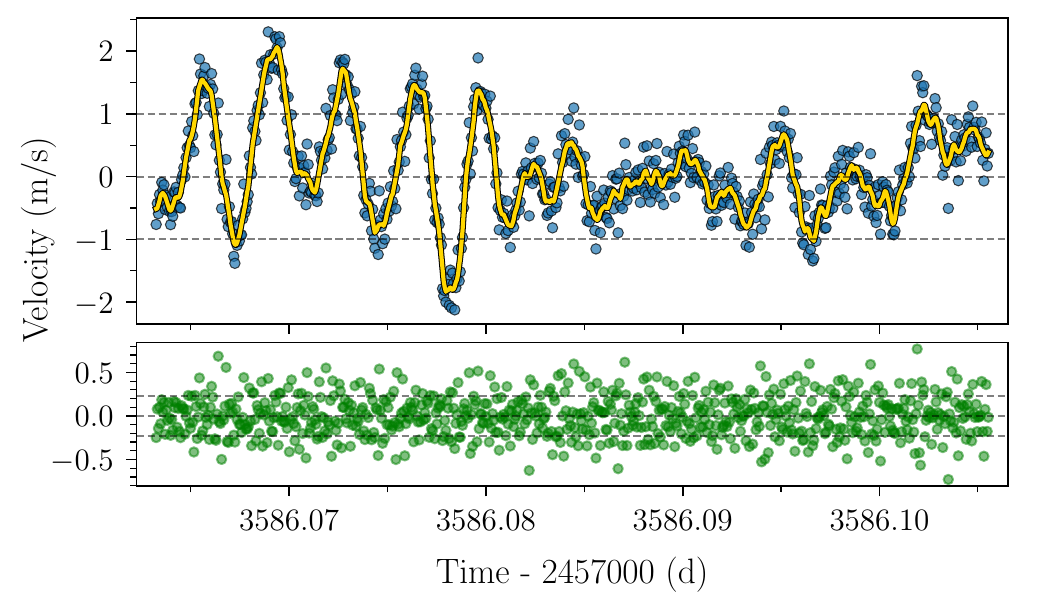}
      \caption{Solar observation test at OT following the change to a QHY600 CMOS detector. Observations were taken over 1 hour in October 2024, resulting in 750 2 s integration measurements with a sampling of 4.88 sec. \textit{Top}: Individual measurements after subtracting a global median (in blue). The yellow curve shows a low-pass-filtered version obtained by applying an Epanechnikov kernel filter with a width of 9 points (${\sim}44$ s). Horizontal lines at $\pm 1$ m/s are added for reference. \textit{Bottom}: Residual after subtracting the filtered version. The horizontal dashed lines indicate the zero residual and the measured scatter (standard deviation) at $\pm 23$ cm/s.}   
         \label{fig:solar_data}
   \end{figure}

A solar altitude-azimuth tracker and scrambling optics were installed for the optical fibre input in 2017 \citep{Andersen2019}. A high-cadence ($4$ s) observing run was conducted over 57 days during 2018, resulting in over 500,\,000 spectra. Due to a malfunction, the active guide servo for the tracker was not in operation, resulting in partially resolved solar observations (\ie not fully Sun-as-a-star observations). The analysis of these data by \citet{Andersen2019} led to the first reference values for the solar global asteroseismic parameters \dnu and \numax from SONG. Importantly, the measured values confirmed the importance of obtaining instrument-specific reference values when using global seismic parameters for solar-like oscillators, such as scaling relations, to assess the impact on \numax of different atmospheric depths probed by different instruments, for instance.

A follow-up analysis of a 30-day subset of the 2018 data by \citet{Breton2022} compared the performance of SONG at mid- to low-frequencies in the $p$-mode regime with observations from other solar observatories. This analysis found that lower-frequency modes could be obtained with Solar-SONG compared to GOLF and BiSON. These results clearly show the potential of high-cadence solar observations with SONG, especially if extended across the full network to improve the duty cycle. 

Following the change of the detector at OT in January 2024 to a QHY600 CMOS with improved sampling from smaller pixels, a 1-hour test was performed in October 2024, gathering 750 spectra at an average sampling interval of 4.88 sec with 2 s integrations. From this test, a precision of ${\sim}23$ cm/s per spectrum was achieved (\fref{fig:solar_data}), corresponding to an improvement in the RV precision of a factor of two. 

We note that the high-cadence Solar-SONG observations described above are not currently accessible via SODA, but can be made available upon request. However, SODA contains an additional 81,\,278 solar spectra that have been obtained at a low cadence since 2014 (see Tables~\ref{tab:song_stars}--\ref{tab:obs}).

At OT, a new initiative called the Magnetometry Unit for SOLar-SONG (MUSOL) also plans to upgrade the fibre feed with a more stable equatorial mount and a spectropolarimetric front-end unit to measure the dipole and quadrupole components of the solar global magnetic field.
The performance of the new detector at OT and the MUSOL initiative show that the prospects for high-quality continuous solar observations from SONG are promising. A working group has recently been formed within the SONG consortium to focus on the Solar-SONG initiative and to consider implementing network-wide solar observations in a steady state.

\section{Using SONG Data}\label{sec:use}
Data products from SONG (\sref{sec:data}) are accessible to members of the SONG community via the SODA database\footnoteref{sodaurl}. We consider anyone who signs up as a SODA user, thereby acknowledging the SONG publication policy and code of conduct, to be a member of the SONG community. 

When using SONG data, the user should first be aware of differences among data files from different detectors and sites, as discussed in \sref{sec:spectra}. 
When working with RV time-series data (\sref{sec:timeseries}) provided via SODA, we note that RVs for all reduced spectra for a given star will be provided. The user is responsible for filtering the data for potential outliers, bad data points, and other issues at a level appropriate to the specific analysis. A very useful quantity in this context is the flux level provided alongside the extracted velocities. Here, it may be prudent to set a lower flux threshold and identify times with significant flux variation---these will often coincide with outliers in the extracted velocities. It is also important to ensure that all data points in a given RV time series have the same exposure time, as this affects the amplitude attenuation of all non-white signals.

\subsection{SODA data access}\label{sec:access}

When searching for data of a specific star, the common practice in SODA is to search by the object name. Generally, the database name resolver will identify the star regardless of the object name entered. However, when observations are not found by searching with the Hipparcos (HIP) ID, for example, it is worthwhile trying other commonly used identifiers, such as the Henry Draper (HD) number. 
To search for observations from multiple stars, it is also possible to search by stellar properties (e.g. sky coordinates, proper motion, and brightness), specific observation time intervals, project IDs, and observatory characteristics (site, iodine or ThAr, and slit).
If the goal is to search for observations of several specific stars, the currently available method is to provide the unique file identifiers for those observations. With this publication, we provide a complete overview of all stars with SONG observations available in SODA through 2025 in \tref{tab:song_stars}, and a full record of all individual observations is available in \tref{tab:obs}. From this, it is possible to extract the file identifiers (File ID) for the observations of interest and retrieve these via SODA.

\subsection{Proposing and using SONG observations}\label{sec:wgs}

Proposals for new observations are principally handled via the SONG working groups (WGs). SONG members are encouraged to interact with other members and the chairs of the different WGs to coordinate which targets are proposed for observations. Users can propose new observations by submitting information detailing the technical aspects of the observations, along with a brief scientific justification, to the SONG science coordinators. All information required for this process is available on SODA. 

Membership of the SONG consortium is available by signing up on SODA. Specific WGs can be joined under the user account profile after approval by the WG chairs and the science coordinator.
Users of SONG data are expected to adhere to the SONG data and publication policies and to properly acknowledge the use of SONG in any published material. These policies and the appropriate acknowledgements are available on SODA. 

\section{Conclusions and outlook}\label{sec:con}

The SONG network has now operated for more than a decade, accumulating more than $580,000$ spectra of 3091 stellar targets between 2014 and 2025 across a broad range of science applications: asteroseismology, stellar variability, binary systems, and exoplanets. This legacy dataset, hosted in the SONG data archive (SODA), is among the most extensive archives of high-resolution time-domain stellar spectroscopy currently available and continues to grow through ongoing observations.

We described SONG's observational capabilities, data products, and pathways for community access. The network remains open to new members, and we encourage the community to make use of the archival observations and the opportunity to propose new observing programs.

A growing emphasis of future SONG observations will be on coordinated campaigns obtained contemporaneously with space-based photometric missions, particularly TESS and the forthcoming PLATO mission. The combination of high-cadence ground-based RV observations with space photometry opens new avenues for joint investigations of stellar oscillations and variability. Such datasets are well placed to constrain oscillation-mode excitation and damping, amplitude ratios, and phase relations between photometric and velocity fluctuations, as already illustrated with the analysis of SONG data for $\beta$ Aql \citep{Kjeldsen2025}. Empirical photometric-RV amplitude ratios derived from these campaigns may also help characterise and ultimately mitigate stellar signals in precision RV exoplanet searches, where stellar activity remains a limiting systematic.

In parallel, the planned network extension with the Lenghu site, along with the recently commissioned APO site, will substantially improve the longitudinal coverage and observing duty cycles. The improved node geometry will reduce diurnal aliasing and enable near-continuous spectroscopic time series. Together, these developments will extend SONG's reach as a leading ground-based facility for long-term stellar time-domain astrophysics.

\section*{Data availability}
Tables~\ref{tab:song_stars} and \ref{tab:obs} are published in their entirety in electronic machine-readable format online at the CDS via anonymous ftp to \url{cdsarc.u-strasbg.fr} ($130.79.128.5$) or via \url{http://cdsweb.u-strasbg.fr/cgi-bin/qcat?J/A+A/}.

\begin{acknowledgements}
The SONG network of telescopes is operated by Aarhus University, Instituto de Astrofísica de Canarias, the National Astronomical Observatories of China, University of Southern Queensland and New Mexico State University. 
We are grateful for the observation and technical support at each site. We especially acknowledge Antonio Pimienta for his role as the curator of the Hertzsprung SONG Telescope at the Observatorio del Teide for the past 15 years, and Jack Okumura for his support of the Mount Kent node. 
Funding for the Stellar Astrophysics Centre was provided by The Danish National Research Foundation (Grant agreement no.: DNRF106). We would like to acknowledge the Villum Foundation, The Danish Council for Independent Research | Natural Science, and the Carlsberg Foundation for the support in building the SONG prototype on Tenerife. 
SONG-Mount Kent was funded in part by an Australian Research Council LIEF grant (LE190100036).
MNL acknowledges support from the ESA PRODEX programme (PEA 4000142995).
Funded/Co-funded by the European Union (ERC, MAGNIFY, Project 101126182). Views and opinions expressed are, however, those of the author(s) only and do not necessarily reflect those of the European Union or the European Research Council. Neither the European Union nor the granting authority can be held responsible for them.
JR acknowledge support from the Carlsberg Foundation under the grant `A four-node SONG network: state-of-the-art opportunities to study stars and exoplanets' and the Australian Research Council through the Laureate Fellowship FL220100117.
RT acknowledges the Head of Department of Physics and the University of Cambridge.
PGB acknowledges support by the Spanish Ministry of Science, Innovation and Universities (MCIN) with the \textit{Ram{\'o}n\,y\,Cajal} fellowship (RYC-2021-033137-I). FG, PLP, PGB, and RAG acknowledge support by the MCIN with the proyecto plan nacional \textit{PLAtoSOnG} (grant no. PID2023-146453NB-I00).
SJM was supported by the Australian Research Council through
Future Fellowship FT210100485.
JL acknowledge the Australian Research Council grant DP210101299.
This research made use of the SONG database SODA (\url{https://soda.phys.au.dk/}), operated and maintained at Aarhus University, DK. 
\end{acknowledgements}

\bibliographystyle{aa} 
\bibliography{biblio} 

\newpage

\begin{appendix}

\onecolumn
\section{Known characteristics}\label{sec:oneyear}

\begin{figure*}[!htbp]
\sidecaption
   \includegraphics[width=0.5\textwidth]{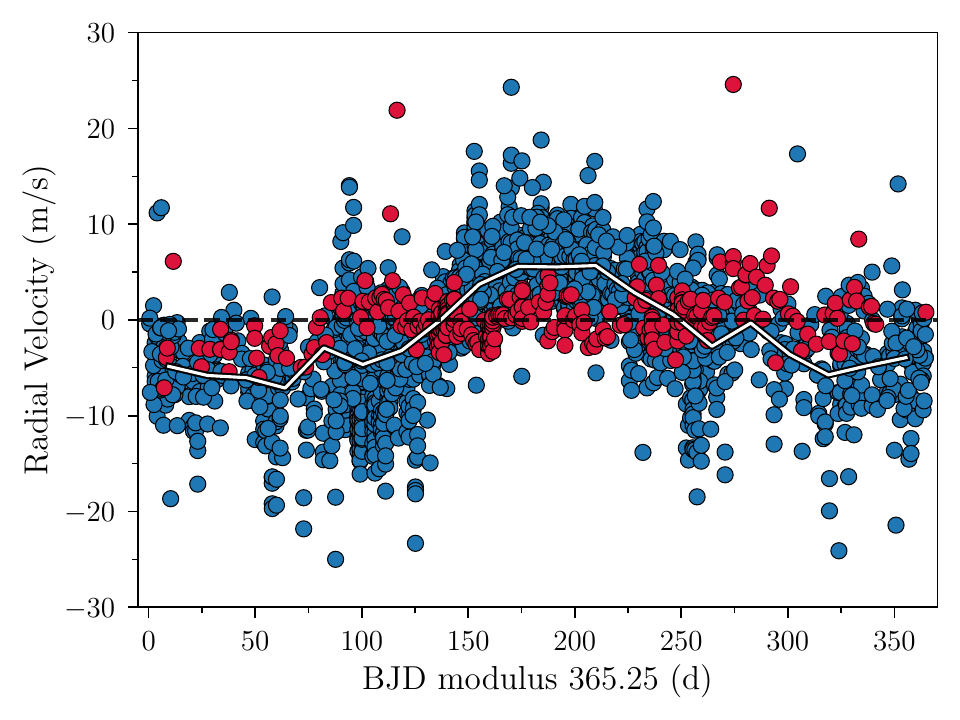}
      \caption{Residual RV variation of the standard star \object{$\sigma$ Dra} phased to a one-year period. Measurements obtained from $2014-2023$ with the Andor detector at OT are shown in blue, while measurements obtained during $2024$ with the new QHY600 detector are shown in red. The white line shows the binned median variation of the Andor measurements. \vspace{4.5cm}
              }
         \label{fig:one_year}
   \end{figure*}

To monitor the spectrograph's RV performance, we began regularly observing RV-stable stars with the iodine cell in early 2014. After approximately one year, it was noticed that \object{$\sigma$ Dra} appeared to exhibit a periodic variation, with a 1-year period. Upon inspection, this phenomenon was seen for several other targets over several years.   
In \fref{fig:one_year} we show the phased residual RV variation of the standard star \object{$\sigma$ Dra}, which illustrates this (blue points). Initially, our attempts to identify the source of the problem focused on the RV extraction code. An independent spectral extraction and RV analysis was kindly performed by Paul Butler and yielded the same RV variation. Likewise, an iodine cell replacement, kindly lent to us by Debra Fischer, did not resolve the problem. This caused us to suspect the Andor detector. In April 2018, the Andor detector was rotated by 90 degrees, but this unfortunately did not resolve the problem. Ultimately, in early 2024, the Andor detector was replaced with a QHY600 detector. 
As seen, the characteristic 1-year periodicity in the RVs from the Andor CCD at OT is effectively removed with the installation of the new QHY600 detector in $2024$. 

\FloatBarrier

\section{Network observing configurations}\label{app:net}
Observing coverage for different network configurations. Note that the limits of the colour scales differ across the network configurations.

\begin{figure}[!htbp]
   \sidecaption
   \includegraphics[width=\textwidth]{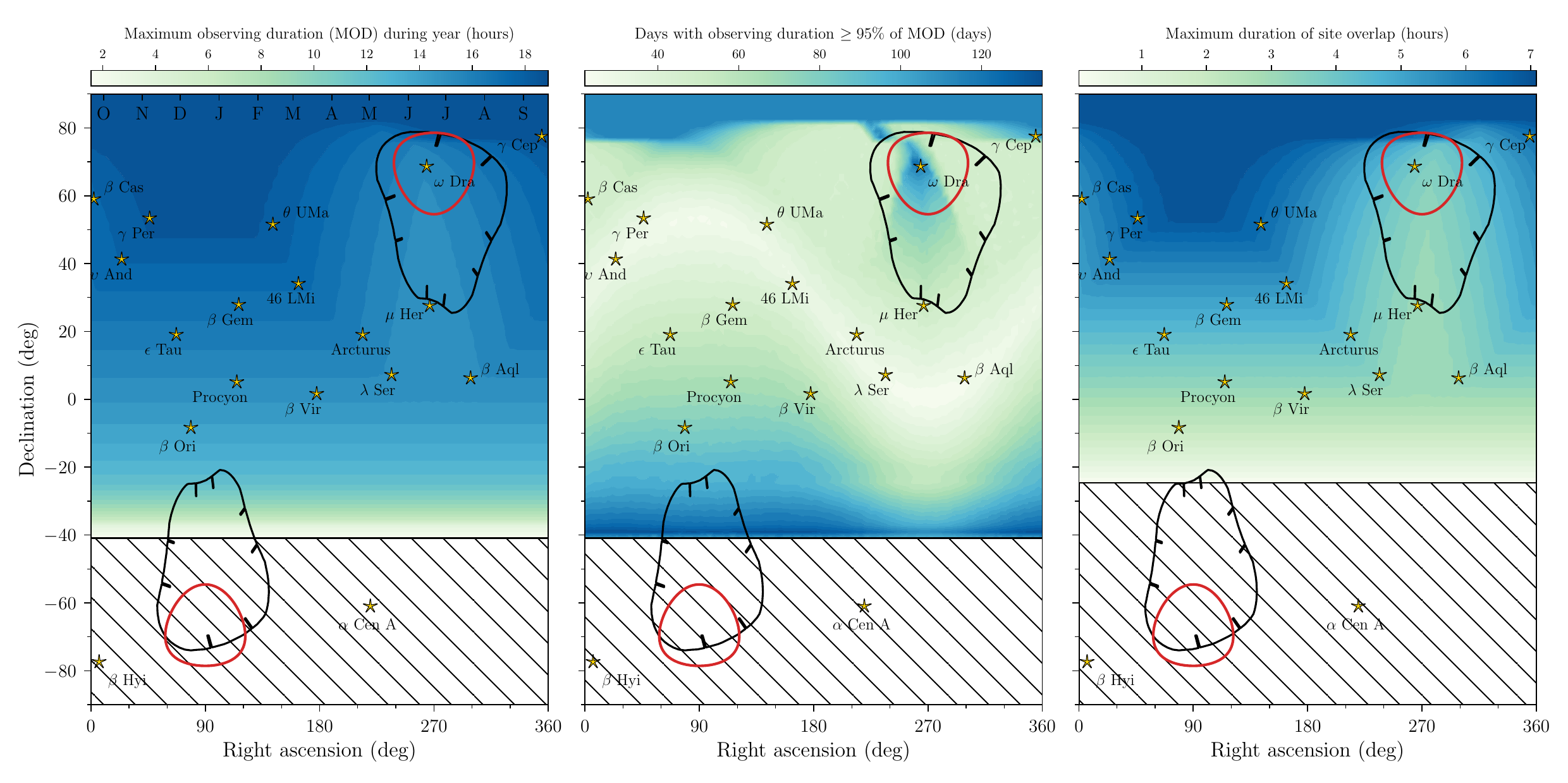}
      \caption{SONG multisite observing coverage, similar to \fref{fig:coverage} but considering only OT and APO.}
         \label{fig:coverage2}
   \end{figure}

\begin{figure}[!htbp]
   \sidecaption
   \includegraphics[width=\textwidth]{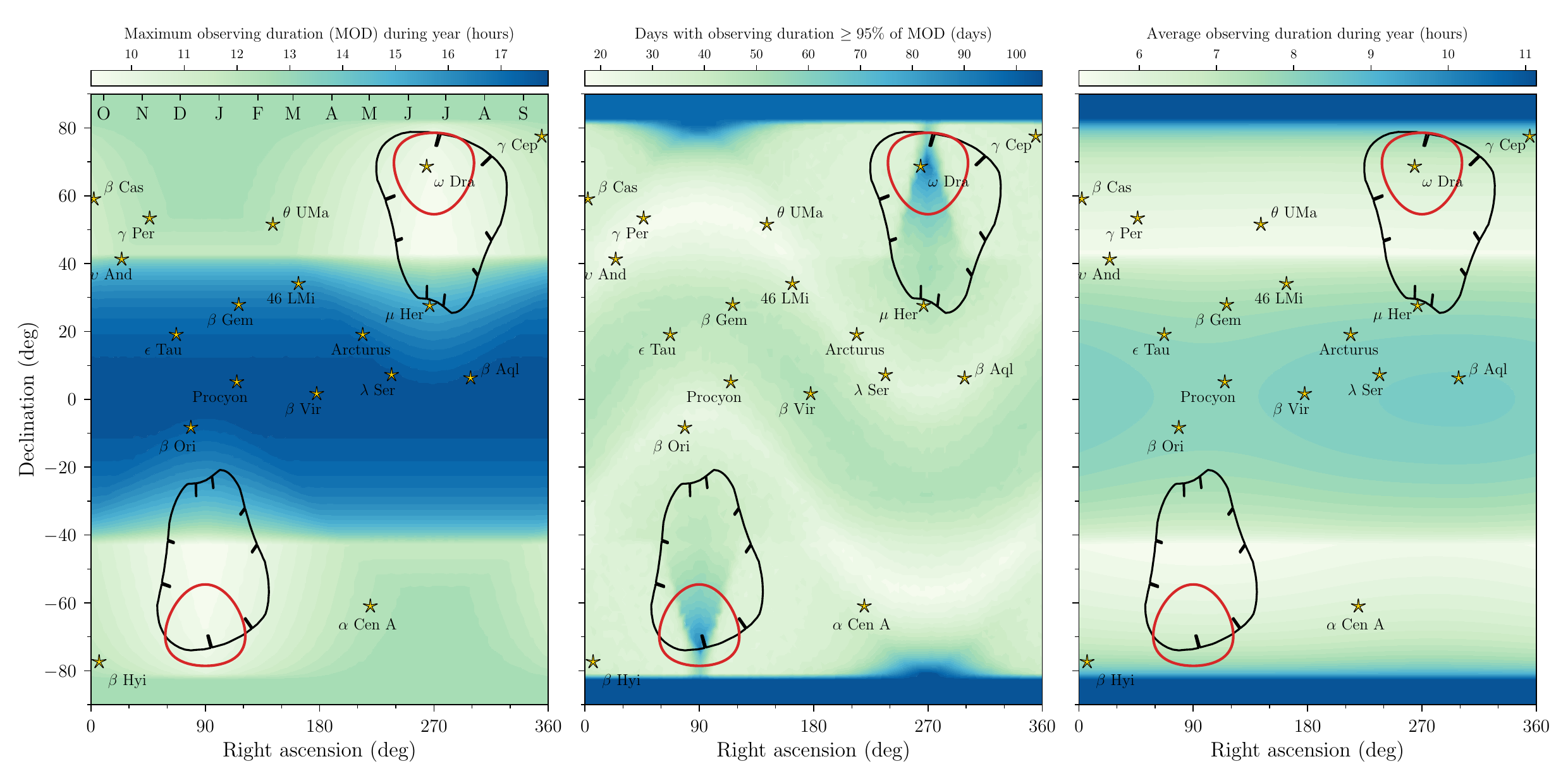}
      \caption{SONG multisite observing coverage, similar to \fref{fig:coverage} but considering only OT and MtK. Given the lack of overlap between OT and MtK, the right panel shows the average observing duration over the year.}
         \label{fig:coverage3}
   \end{figure}

\begin{figure}[!htbp]
   \sidecaption
   \includegraphics[width=\textwidth]{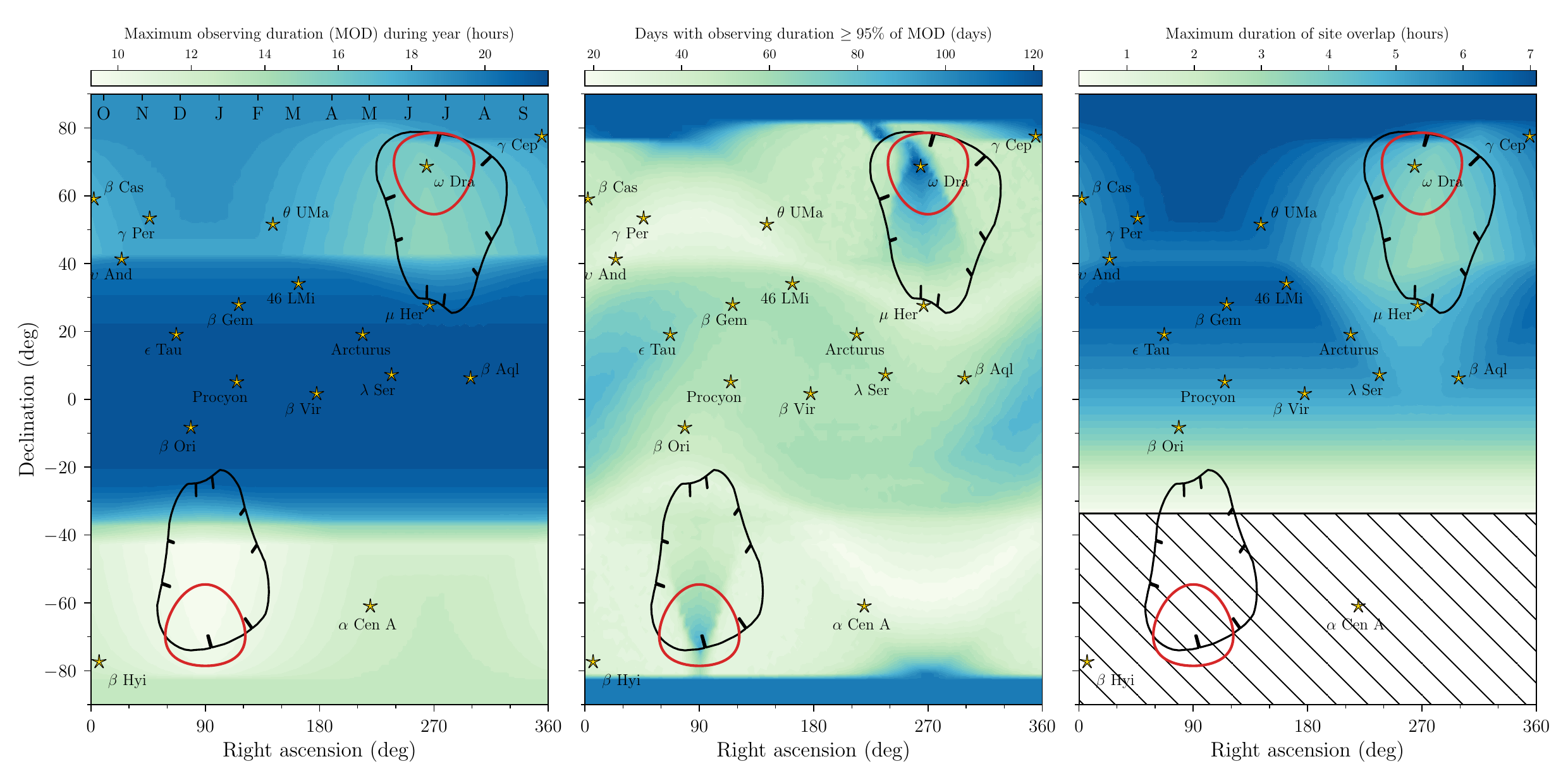}
      \caption{SONG multisite observing coverage, similar to \fref{fig:coverage} but considering only OT, MtK, and APO.}
         \label{fig:coverage4}
   \end{figure}
\clearpage

\section{SONG target and observations tables}
\FloatBarrier
\setlength\tabcolsep{0pt} 
\begin{table*}[htbp!] 
\centering 
\caption{Stars observed by SONG.} 
\label{tab:song_stars}
\scriptsize 
\renewcommand{\arraystretch}{0.8}  
\begin{tabular*}{\linewidth}{@{\extracolsep{\fill}}llllllccccc@{}} 
\toprule 
Name & SODA ID & IAU name & TIC & HIP & HD & $V$ & RA & Dec & No. Spec. & Ratio\\ 
  &      & &  &  & & (mag) & (deg) & (deg) & & (Iodine/ThAr in \%) \\ 
\midrule 
$\mu$$^1$ Her & 41 &   & 460067868 & 86974 & 161797 &3.42 & $266.6$ & $27.7$ & 127165 & 80/20\\   
$\alpha$ CMi & 167 & Procyon & 280310048 & 37279 & 61421 &0.37 & $114.8$ & $5.2$ & 75902 & 90/10\\   
$\gamma$ Cep & 223 & Errai & 367912480 & 116727 & 222404 &3.22 & $354.8$ & $77.6$ & 39083 & 50/50\\   
$\beta$ Aql & 691 & Alshain & 375621179 & 98036 & 188512 &3.71 & $298.8$ & $6.4$ & 35797 & 100/0\\   
$\beta$ Vir & 27 & Zavijava & 366661076 & 57757 & 102870 &3.60 & $177.7$ & $1.8$ & 28412 & 100/0\\   
$\alpha$ Cen A & 339 & Rigil Kentaurus & 471011145 & 71683 &   &0.01 & $219.9$ & $-60.8$ & 20787 & 90/10\\  
$\alpha$ Boo & 161 & Arcturus & 459832522 & 69673 & 124897 &-0.05 & $213.9$ & $19.2$ & 20417 & 90/10\\   
$\theta$ UMa & 355 &   & 150226696 & 46853 & 82328 &3.18 & $143.2$ & $51.7$ & 11085 & 100/0\\   
$\beta$ Hyi & 3123 &   & 267211065 & 46853 & 2151 &2.79 & $6.4$ & $-77.3$ & 10205 & 100/0\\   
$\gamma$ Psc & 720 &   & 262691761 & 114971 & 219615 &3.70 & $349.3$ & $3.3$ & 10052 & 100/0\\   
$\epsilon$ Tau & 130 & Ain & 17554529 & 20889 & 28305 &3.53 & $67.2$ & $19.2$ & 8453 & 90/10\\   
$\delta$ Eri & 407 & Rana & 38511251 & 17378 & 23249 &3.54 & $55.8$ & $-9.8$ & 8234 & 100/0\\   
$\sigma$ Dra & 4 & Alsafi & 259237827 & 96100 & 185144 &4.68 & $293.1$ & $69.7$ & 7788 & 20/80\\   
$\beta$ Gem & 8 & Pollux & 423088367 & 37826 & 62509 &1.14 & $116.3$ & $28.0$ & 7764 & 100/0\\   
$\iota$ Per & 116 &   & 116988032 & 14632 & 19373 &4.05 & $47.3$ & $49.6$ & 7729 & 100/0\\   
$\theta$ Cen & 2823 & Menkent & 179323446 & 68933 & 123139 &2.05 & $211.7$ & $-36.4$ & 7669 & 100/0\\   
$\lambda$ Ori A & 310 & Meissa & 436103278$^{\dagger}$ & 26207$^{\dagger}$ & 36861 &3.47 & $83.8$ & $9.9$ & 6939 & 0/100\\   
$\gamma$ Per & 124 &   & 116553919 & 14328 & 18925 &2.93 & $46.2$ & $53.5$ & 6731 & 80/20\\   
$\eta$ Cas & 176 & Achird & 445258206 & 3821 & 4614 &3.44 & $12.3$ & $57.8$ & 6058 & 80/20\\   
$35$ Dra & 2335 &   & 441813918 & 87234 & 163989 &5.04 & $267.4$ & $77.0$ & 5978 & 100/0\\   
$\epsilon$ Ori & 139 & Alnilam & 427451176 & 26311 & 37128 &1.69 & $84.1$ & $-1.2$ & 5585 & 0/100\\   
$\epsilon$ Cyg & 2 & Aljanah & 232853959 & 102488 & 197989 &2.48 & $311.6$ & $34.0$ & 5373 & 100/0\\   
$\upsilon$ And & 284 & Titawin & 189576919 & 7513 & 9826 &4.10 & $24.2$ & $41.4$ & 4953 & 90/10\\   
$46$ LMi & 165 & Praecipua & 85321185 & 53229 & 94264 &3.83 & $163.3$ & $34.2$ & 4576 & 20/80\\   
$\beta$ Ori & 143 & Rigel & 231308237 & 24436 & 34085 &0.13 & $78.6$ & $-8.2$ & 4446 & 0/100\\   
$\lambda$ Sco & 3121 & Shaula & 465088681 & 24436 & 158926 &1.63 & $263.4$ & $-37.1$ & 4296 & 30/70\\   
$\eta$ Cep & 698 &   & 372682437 & 102422 & 198149 &3.41 & $311.3$ & $61.8$ & 3885 & 100/0\\   
$\xi$ Hya & 2818 &   & 57570218 & 56343 & 100407 &3.54 & $173.2$ & $-31.9$ & 3677 & 100/0\\   
$\alpha$ Tau & 164 & Aldebaran & 245873777 & 21421 & 29139 &0.86 & $69.0$ & $16.5$ & 3388 & 80/20\\   
$\theta$$^1$ Tau & 479 &   & 245792850 & 20885 & 28307 &3.84 & $67.1$ & $16.0$ & 2876 & 50/50\\   
$\beta$ CVn & 28 & Chara & 458445966 & 61317 & 109358 &4.25 & $188.4$ & $41.4$ & 2780 & 40/60\\   
\bottomrule
\end{tabular*} 
\tablefoot{\tiny Table~\ref{tab:song_stars} is published in its entirety in the machine-readable format at the CDS via anonymous ftp to \url{cdsarc.u-strasbg.fr} ($130.79.128.5$) or via \url{http://cdsweb.u-strasbg.fr/cgi-bin/qcat?J/A+A/}. A portion is shown here for guidance regarding its form and content. The table provides information on the stars observed by SONG, sorted according to the total number of SONG spectra available (`No. Spec'). The first six columns provide stellar identifiers: `Name', giving either the Bayer/Flamsteed designation, the primary identifier according to \texttt{SIMBAD}, or, in some cases, the identifier from the General Catalogue of Variable Stars \citep[GCVS;][]{Samus2017}; `SODA ID', the unique identifier used in the SODA database (and referenced in Table~\ref{tab:obs}); `IAU name' giving the proper name adopted by the IAU Working Group on Star Names (\url{https://exopla.net/star-names/modern-iau-star-names/}); together with the TESS (`TIC'), Hipparcos (`HIP'), and Henry Draper (`HD') identifiers. `$V$' gives the visual magnitude, and `Ratio' the percentage split of the `No. Spec' spectra between iodine and ThAr observations. \tablefoottext{$\dagger$}{Identifier or name refers to the system, rather than the individual star.}}
\end{table*}

\setlength\tabcolsep{0pt} 
\begin{table*} 
\centering 
\caption{Individual observations made by SONG.} 
\label{tab:obs}
\small 
\renewcommand{\arraystretch}{0.8}  
\begin{tabular*}{\linewidth}{@{\extracolsep{\fill}}llccccc@{}} 
\toprule 
SODA ID & File ID & Site & Type & Slit & Exp. time & Obs. beg.\\ 
  &      & (OT; MtK; Del) & (Iodine; ThAr) &  & (s) & (MJD) \\ 
\midrule 
 143 & 296326 & OT & ThAr & 6 &15 & 57012.1198951 \\   
 8 & 1146423 & OT & Iodine & 6 &40 & 59952.2441365 \\   
 167 & 1356560 & MtK & Iodine & - &60 & 59986.4840528 \\   
 41 & 369909 & OT & ThAr & 6 &120 & 56799.0840766 \\   
 41 & 1054322 & OT & Iodine & 6 &120 & 59747.9668116 \\   
 167 & 622512 & OT & Iodine & 8 &20 & 58508.1668259 \\   
 167 & 630696 & OT & Iodine & 8 &20 & 58511.1326576 \\   
 165 & 388035 & OT & ThAr & 6 &240 & 56728.9533241 \\   
 8 & 94232 & OT & Iodine & 5 &20 & 57391.8831069 \\   
 407 & 562955 & OT & Iodine & 6 &120 & 58409.9682903 \\   
\bottomrule
\end{tabular*} 
\tablefoot{\tiny Table~\ref{tab:obs} is published in its entirety in the machine-readable format at the CDS     via anonymous ftp to \url{cdsarc.u-strasbg.fr} ($130.79.128.5$) or via \url{http://cdsweb.u-strasbg.fr/cgi-bin/qcat?J/A+A/}.     A portion is shown here for guidance regarding its form and content.     The table provides information on the individual observations made by SONG, sorted according to the unique identifier `SODA ID' in the SODA database for the star observed.     The star name corresponding to a given SODA ID can be obtained from Table~\ref{tab:song_stars}. In the example shown, the entries have been randomly picked from the full table. `File ID' gives the identifier for the specific observation, enabling retrieval of the file from the SODA database. The site of the observation is provided by `Site', including     OT, MtK, and Del (Delingha) for observations through 2025. The `Type' indicates if the observation was made in the iodine or ThAr mode, `Slit' gives the adopted slit (not included for fibre-based observations from MtK), the exposure time is given with `Exp. time', while `Obs. beg.' gives the time at the beginning of the observation.     }
\end{table*} 

\end{appendix}
\end{document}